\documentclass[twocolumn]{autart}    

\usepackage{mysty}
\usepackage{bbold,amsmath,amsfonts,amssymb}
\usepackage{url}
\allowdisplaybreaks
\usepackage{environ}
\usepackage{cite}
\usepackage{accents}
\usepackage{balance}
\usepackage{ifthen}
\newboolean{showcomments}
\setboolean{showcomments}{true}
\usepackage[usenames,dvipsnames]{xcolor}
\usepackage{todonotes}

\definecolor{bleudefrance}{rgb}{0.19, 0.55, 0.91}
\definecolor{ao(english)}{rgb}{0.0, 0.5, 0.0}

\newcommand{\addcite}[0]{\ifthenelse{\boolean{showcomments}}
{\textcolor{purple}{(add cite(s)) }}{}}%

\newcommand{\enrique}[1]{  \ifthenelse{\boolean{showcomments}}
{\todo[inline,color=bleudefrance]{Enrique: #1}}{}}
\newcommand{\emmargin}[1]{\ifthenelse{\boolean{showcomments}}{\marginpar{\color{bleudefrance}\tiny EM: #1}}{}}
\newcommand{\hancheng}[1]{  \ifthenelse{\boolean{showcomments}}
{\todo[inline,color=orange]{Hancheng: #1}}{}}

\newboolean{showedits}
\setboolean{showedits}{true}
\usepackage[xcolor=bleudefrance]{changes}
\definechangesauthor[color=bleudefrance]{EM}
\newcommand{\aem}[1]{
\ifthenelse{\boolean{showedits}}
{\added[id=EM]{#1}}
{#1}
}
\newcommand{\chem}[2]{
\ifthenelse{\boolean{showedits}}
{\replaced[id=EM]{#1}{#2}}
{#1}
}
\newcommand{\dem}[1]{
\ifthenelse{\boolean{showedits}}
{\deleted[id=EM]{#1}}
{}
}
\newcommand{\hl}[1]{
\ifthenelse{\boolean{showedits}}
{{\color{bleudefrance}#1}}
{#1}
}

\setboolean{showedits}{true}
\setboolean{showcomments}{true}
\newboolean{archive}
\setboolean{archive}{false}
\newboolean{bio}
\setboolean{bio}{true}
\newboolean{color}
\setboolean{color}{false}
\usepackage{verbatim}
\usepackage{graphicx}          
\newif\ifshownotes
\shownotestrue
\definecolor{notetext}{rgb}{0.7,0,0}

\begin{document}

\begin{frontmatter}
\title{A Frequency Domain Analysis of Slow Coherency in Networked Systems}

\thanks[footnoteinfo]{Preliminary version of this work, covering an alternative version of the results in Section \ref{sec:dymC}, was presented in \cite{min2019cdc}.}

\author[jhu]{Hancheng Min}\ead{hanchmin@jhu.edu},    
\author[lund]{Richard Pates}\ead{richard.pates@control.lth.se},               
\author[jhu]{Enrique Mallada}\ead{mallada@jhu.edu}  

\address[jhu]{Johns Hopkins University, Baltimore, MD, U.S.A.}  
\address[lund]{Lund University, Lund, Sweden}             


\begin{abstract}                          
Network coherence generally refers to the emergence of simple aggregated dynamical behaviours, despite heterogeneity in the dynamics of the subsystems that constitute the network.
In this paper, we develop a general frequency domain framework to analyze and quantify the level of network coherence that a system exhibits by relating coherence with a low-rank property of the system's input-output response. More precisely, for a networked system with linear dynamics and coupling, we show that, as the network's \emph{effective algebraic connectivity} grows, the system transfer matrix converges to a rank-one transfer matrix representing the coherent behavior. Interestingly, the non-zero eigenvalue of such a rank-one matrix is given by the harmonic mean of individual nodal dynamics, and we refer to it as the coherent dynamics. Our analysis unveils the frequency-dependent nature of coherence and a non-trivial interplay between dynamics and network topology. 
We further show that many networked systems can exhibit similar coherent behavior by establishing a concentration result in a setting with randomly chosen individual nodal dynamics.
\end{abstract}

\end{frontmatter}

\section{Introduction}\label{sec:intro}
The study of coordinated behavior in network systems has been a popular subject of research in many fields, including physics~\cite{Bressloff1999}, chemistry~\cite{Kiss2002}, social sciences~\cite{DeGroot1974}, and biology~\cite{Mirollo1990}. Within engineering, coordination is essential for the proper operation of many networked systems, including power networks~\cite{jpm2017cdc,Paganini2019tac}, data and sensor networks~\cite{mmhzt2015ton,m2014phd-thesis}, and autonomous transportation~\cite{Sepulchre2008706,Olfati-Saber2007,Jadbabaie2003988,Bamieh2012}. Among many forms of coordination, \emph{coherence} refers to the ability of a group of nodes to have a similar dynamic response to some external disturbance~\cite{Chow2013}. While coherence analysis is useful in understanding the collective behavior of large networks, little do we know about the underlying mechanism that causes such coherent behavior to emerge in various networks.

Classic slow coherency analyses~\cite{chow1982time,ramaswamy1996,romeres2013,tyuryukanov2021,fritzsch2022} (with applications mostly to power networks) usually consider the second-order electro-mechanical model without damping: $\ddot{x}=-M^{-1}Lx$, where $M$ is the diagonal matrix of machine inertias, and $L$ is the Laplacian matrix whose elements are synchronizing coefficients between pair of machines. The coherency or synchrony~\cite{ramaswamy1996} (a generalized notion of coherency) is identified by studying the first few slowest eigenmodes (eigenvectors with small eigenvalues) of $M^{-1}L$. The analysis can be carried over to the case of uniform~\cite{chow1982time} and non-uniform~\cite{romeres2013} damping. However, such state-space-based analysis is limited to very specific node dynamics (second order) and does not account for more complex dynamics or controllers that are usually present at a node level; e.g., in the power systems literature~\cite{jpm2021tac, jbvm2021lcss, ekomwenrenren2021}. Moreover, it is widely known that such coherence is related to strong interconnection among the nodes, such relation is not formally justified in the aforementioned slow coherency analyses.

A vast body of work, triggered by the seminal paper~\cite{Bamieh2012}, has quantitatively studied the role of the network topology in the emergence of coherence. Examples include, directed~\cite{Tegling19} and undirected~\cite{Oral17} consensus networks, transportation networks~\cite{Bamieh2012}, and power networks~\cite{Paganini2019tac,Bamieh2013,Andreasson17,psf17cdc}. 
The key technical approach amounts to quantify the level of coherence by computing the $\mathcal{H}_2$-norm of the system for appropriately defined nodal disturbance and performance signals. Broadly speaking, the analysis shows a reciprocal dependence between the performance metrics and the non-zero eigenvalues of the network graph Laplacian, validating the fact that strong network coherence (low $\mathcal{H}_2$-norm) results from the high connectivity of the network (large Laplacian eigenvalues).
Unfortunately, the analysis strongly relies on a homogeneity~\cite{Bamieh2012,Tegling19, Oral17,Bamieh2013, Andreasson17,psf17cdc} or proportionality~\cite{Paganini2019tac} assumption of the nodal transfer functions, and thus fails to characterize how individual heterogeneous node dynamics affect the overall coherent network response.
\subsection{Our contribution}
{\color{black}In this paper, we seek to overcome these limitations by formalizing network coherence through a low-rank structure of the system transfer matrix that appears when the network feedback gain is high. 
This frequency domain analysis provides a deeper characterization of the role of both, network topology and node dynamics, on the coherent behavior of the network. In particular, our results make substantial contributions towards the understanding of coordinated and coherent behavior of network systems in many ways:
\begin{itemize}
    \item We present a general framework in the frequency domain to analyze the coherence of heterogeneous networks. We show that network coherence emerges as a low-rank structure of the system transfer matrix as we increase the effective algebraic connectivity--a frequency-varying quantity that depends on the network coupling strength and  dynamics.
    \item Our analysis applies to networks with  heterogeneous nodal dynamics, and further provides an explicit characterization in the frequency domain of the coherent response to disturbances as the harmonic mean of individual nodal dynamics. Thus, in this way, our results highlight the contribution of individual nodal dynamics to the network's coherent behavior.
    \item We formally connect our frequency-domain results with explicit time-domain $L_\infty$ bounds on the difference between individual nodal responses and the coherent dynamic response to certain classes of input signals, suggesting that network coherence is a frequency-dependent phenomenon. That is, the ability of nodes to respond coherently depends on the frequency composition of the input disturbance.  
    \item By providing an exact characterization of the network's coherent dynamics, our analysis can be further applied in settings where only distributional information of the network composition is known. More precisely, we show that the coherent dynamics of tightly-connected networks with possibly random nodal dynamics are well approximated by a deterministic transfer function that only depends on the statistical distribution of node dynamics. 
\end{itemize}}

Notably, the problem of characterizing coherent dynamic response is unique to heterogeneous networks since the coherent dynamics for homogeneous networks are exactly equal to the common nodal dynamics. In real applications, however,  such as power networks, such characterization is relevant to model reduction~\cite{Germond1978} and control design~\cite{jbvm2021lcss}. Our analysis provides, in the asymptotic sense, the exact characterization of coherent dynamics that can be used in control design for heterogeneous networks.
\subsection{Other related work}

\textbf{Consensus and synchronization}:
Consensus~\cite{DeGroot1974, Olfati-Saber2007, Jadbabaie2003988, Bamieh2012, Tegling19,Olfati-Saber20041520, Ghaedsharaf2019} refers to the ability of the network nodes to asymptotically reach a common value over some quantities of interest. 
Synchronization~\cite{Mirollo1990,mmhzt2015ton,m2014phd-thesis,Sepulchre2008706,Nair2008661,Kim2011200,Wieland2011} refers to the ability of network nodes to follow a commonly defined trajectory. Although for nonlinear systems synchronization is a structurally stable phenomenon, in the linear case~\cite{Nair2008661, Sepulchre2008706, Kim2011200, Wieland2011}, synchronization requires the existence of a common internal model that acts as a virtual leader~\cite{Kim2011200,Wieland2011}. As such, consensus and synchronization are coordinated behavior generally achieved in steady state, and requires a common internal model for every node. On the contrary, the network can exhibit coherent behavior during transient phase (a formal comparison is presented in Section \ref{ssec_comp_time}), and coherence exists even without a common internal model. 

\textbf{Area aggregation and dynamic equivalents}:
For a group of nodes that exhibit coherent behavior, one can construct dynamic equivalents~\cite{chow1982time,ramaswamy1996} that characterize the slow coherence. Finding the dynamic equivalent, or an aggregate model, for interconnected power generators is long standing research subject in power system literature. Previously proposed aggregation model~\cite{Germond1978,Anderson1990,romeres2013,Guggilam2018,Paganini2019tac}, mostly assume first- or second-order generator dynamics, which does not account for more complex dynamics or controllers~\cite{jpm2021tac, jbvm2021lcss, ekomwenrenren2021}. Our explicit characterization of coherent dynamics provides a principled way to obtain an aggregate model for general node dynamics.

\subsection{Paper organization}
The paper is organized as follows. In Section \ref{sec:ptw_conv} we discuss the network coherence as a low-rank property of the network transfer matrix. In Section \ref{sec_to_time}, we discuss the time-domain implication of such coherence in transfer matrix. In Section \ref{sec:dymC}, the dynamics concentration in large-scale networks is discussed. In Section \ref{sec:exmples}, we apply our analysis to synchronous generator networks. Lastly we conclude with a discussion on future research in Section \ref{sec:conclusion}. 

\emph{Notation:}~For a vector $x$, $\|x\|=\sqrt{x^\top x}$ denotes the $2$-norm of $x$, and for a matrix $A$, $\sigma_{\min}(A)$ denotes the minimum singular value of $A$, $\|A\|$ denotes the spectral norm of $A$.  Particularly, if $A$ is real symmetric, we let $\lambda_i(A)$ denote the $i$th smallest eigenvalue of $A$. 
We let $\dg\{x_i\}_{i=1}^n$ denote a $n\times n$ diagonal matrix with diagonal entries $x_i$.
We let $I_n$ denote the identity matrix of order $n$, $\one$ denote column vector $[1,\cdots,1]^\top  $, $[n]$ denote the set $\{1,2,\cdots,n\}$ and $\mathbb{N}_+$ denote the set of positive integers. Also, we write complex numbers as $a+jb$, where $j=\sqrt{-1}$. We denote $\mathbb{C}$ the field of complex number, and define the following subsets $\mathcal{B}(s_0,\delta):=\{s\in\mathbb{C}:|s-s_0|\leq \delta\}$.

\section{Problem Setup}\label{sec:prem}
\begin{figure}[ht]
    \centering
	\includegraphics[height=2.5cm]{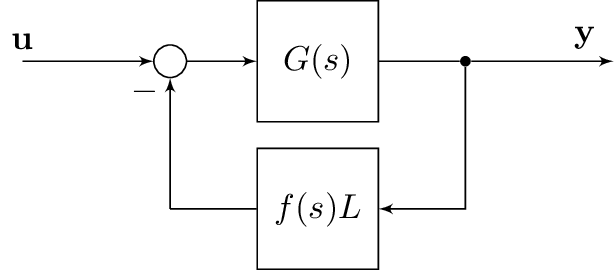}
	\caption{Block diagram of networked dynamical systems}\label{blk_p_n}
\end{figure}

Consider a network consisting of $n$ nodes ($n\geq 2$), indexed by $i\in[n]$ with the block diagram structure in Fig.\ref{blk_p_n}. $L$ is the Laplacian matrix of the weighted graph that describes the network interconnection. We further use $f(s)$ to denote the transfer function representing the dynamics of network coupling, and $G(s)=\mathrm{diag}\{g_i(s)\}$ to denote the nodal dynamics, with $g_i(s),\ i\in[n]$, being an SISO transfer function representing the dynamics of node $i$. Throughout this paper, we assume all $g_i(s),\ i=1,\cdots,n$ and $f(s)$ are rational proper transfer functions, and the Laplacian matrix $L$ is real symmetric.

Under this setting, we can compactly express the transfer matrix from the input signal vector $\mathbf{u}$ to the output signal vector $\mathbf{y}$ by
\begin{align}
    T(s)&=\; (I_n+G(s)f(s)L)^{-1}G(s)\nonumber\\
    &=\; (I_n+\mathrm{diag}\{g_i(s)\}f(s)L)^{-1}\mathrm{diag}\{g_i(s)\}\,.\label{eq_T_explict}
\end{align}
Many existing networks can be represented by this structure. For example, for the first-order consensus network~\cite{Olfati-Saber20041520,Olfati-Saber2007}, $f(s)=1$, and the node dynamics are given by $g_i(s)=\frac{1}{s}$. For power networks~\cite{Andreasson17,Paganini2019tac}, $f(s)=\frac{1}{s}$, $g_i(s)$ are the dynamics of the generators, and $L$ is the Laplacian  matrix representing the sensitivity of power injection w.r.t. bus phase angles. Finally, in transportation networks~\cite{Jadbabaie2003988,Olfati-Saber2007}, $g_i(s)$ represent the vehicle dynamics whereas $f(s)L$ describes local inter-vehicle information transfer. 

Since $L$ has an eigendecomposition $L=V\Lambda V^\top $ where $V=\lhp\frac{\one}{\sqrt{n}},V_\perp\rhp$, $VV^\top =V^\top V=I_n$, and $\Lambda=\mathrm{diag}\{\lambda_i(L)\}$ with $0=\lambda_1(L)\leq\lambda_2(L)\leq \cdots\leq \lambda_n(L)$, we can rewrite $T(s)$ as
\begin{align}
    T(s)&=\; (I_n+\mathrm{diag}\{g_i(s)\}f(s)L)^{-1}\mathrm{diag}\{g_i(s)\}\nonumber\\
    &=\; (\mathrm{diag}\{g^{-1}_i(s)\}+f(s)L)^{-1}\nonumber\\
    &=\; (\mathrm{diag}\{g^{-1}_i(s)\}+f(s)V\Lambda V^\top )^{-1}\nonumber\\
    &=\; V(V^\top \mathrm{diag}\{g^{-1}_i(s)\}V+f(s)\Lambda)^{-1}V^\top \,.\label{eq_T_eigenform}
\end{align}
As we mentioned in the introduction, we are interested in the regime where the closed-loop system $T(s)$ of \eqref{eq_T_explict} has a low-rank structure. To gain some insight, we first consider the following simplified example.
\subsection{Motivating example: homogeneous network}
{\ifthenelse{\boolean{color}}{\color{blue}}{}Suppose $g_i(s)$ are homogeneous, i.e., $g_i(s)=g(s)$. Then using \eqref{eq_T_eigenform} one can decompose $T(s)$ as follows
\be
    T(s) \!=\! \frac{1}{n}g(s)\one\one^\top \!+V_\perp\dg\lb \frac{1}{g^{-1}(s)\!+\!f(s)\lambda_i(L)}\rb_{i=2}^n \!\!\!V_\perp^\top \,,\label{eq_T_homo_decomp}
\ee
where the network dynamics decouple into two terms: 1) the dynamics $\frac{1}{n}g(s)\one\one^\top $ that is independent of network topology and corresponds to the coherent behavior of the system; 2) the remaining dynamics that are dependent on the network structure via both, the eigenvalues $\lambda_i(L), i=2,\cdots,n$ and the eigenvectors $V_\perp$. 
Notice that $|f(s)\lambda_2(L)|\leq|f(s)\lambda_i(L)|, i=2,\dots,n$, then $\frac{1}{n}g(s)\one\one^\top $ is dominant in $T(s)$ as long as $|f(s)\lambda_2(L)|$ (later referred as \emph{effective algbraic connectivity}),  is large enough to make the norm of the second term in \eqref{eq_T_homo_decomp} sufficiently small. Following such observation, we can find two regimes where the coherent dynamics $\frac{1}{n}g(s)\one\one^\top $ is dominant:
\begin{enumerate}
    \item (\emph{High network connectivity}) 
    If a compact set $S\subset \compl$ contains neither zeros nor poles of $g(s)$, then we have
    $
        \lim_{\lambda_2(L)\ra \infty}\sup_{s\in S}\lV T(s)-\frac{1}{n}g(s)\one\one^\top \rV=0\,.
    $
    \item (\emph{High gain in coupling dynamics}) If $s_0$ is a pole of $f(s)$, and the network is connected, i.e., $\lambda_2(L)> 0$, then we have
    $\lim_{s\ra s_0}\lV T(s)-\frac{1}{n}g(s)\one\one^\top \rV=0\,.$
\end{enumerate}
Such convergence results suggest that if 1) the network has high algebraic connectivity, or 2) our point of interest in frequency domain is close to pole of $f(s)$, the response of the entire system is close to one of $\frac{1}{n}g(s)\one\one^\top $. We refer $\frac{1}{n}g(s)\one\one^\top $ as the coherent dynamics\footnote{We also refer $g(s)$ as the coherent dynamics since transfer matrix of the form $\frac{1}{n}g(s)\one\one^\top $ is uniquely determined by its non-zero eigenvalue $g(s)$.} in the sense that in such system, the inputs are aggregated, and all nodes have exactly the same response to the aggregate input.
\emph{Therefore, coherence of the network corresponds, in the frequency domain, to  the property that the network's transfer matrix approximately having a particular rank-one structure}. 

The aforementioned analysis can be extended to the case with proportionality assumption, i.e., $g_i(s)=p_ig(s)$ for some $g(s)$ and $p_i>0,i=1,\cdots,n$, where one can still obtain decoupled dynamics through proper coordinate transformation~\cite{Paganini2019tac} and the coherent dynamics are again characterized by the common dynamics $g(s)$. However, it is challenging to analyze the transfer matrix $T(s)$  without the proportionality assumption: First, it is unclear whether low-rank structure would even emerge under high network connectivity or high gain in the coupling dynamics; Then most importantly, there is no obvious choice for coherent dynamics, hence characterizing the coherent dynamics is a non-trivial problem unique to heterogeneous networks, and no existing work has shown an explicit characterization.

}
\subsection{Goal of this work}\label{ssec:goal}
Our work precisely aims at understanding the coherent dynamics of non-proportional heterogeneous networks. We would like to show that even when $g_i(s)$ are heterogeneous, similar results as in the motivating example still hold. More precisely, we show that, in Section \ref{sec:ptw_conv}, $T(s)$ converges to a rank-one transfer matrix of the form $\frac{1}{n}\bar{g}(s)\one\one^\top $, as the effective algebraic connectivity $|f(s)\lambda_2(L)|$ increases. 
However, unlike the homogeneous node dynamics case where the coherent behavior is driven by $\bar g(s)=g(s)$, the coherent dynamics $\bar{g}(s)$ are given by the harmonic mean of $g_i(s),i=1,\cdots,n$, i.e.,
\be
    \bar{g}(s)=\lp \frac{1}{n}\sum_{i=1}^ng_i^{-1}(s)\rp^{-1}\,.\label{eq_g_bar}
\ee
The convergence results are presented in the aforementioned two regimes: high network connectivity and high gain in coupling dynamics. We then discuss in Section \ref{sec_to_time} their implications on network's time-domain response: 
\begin{enumerate}
    \item Network with high connectivity responds coherently to a wide class of input signals;
    \item Network with coupling dynamics $f(s)=\frac{1}{s}$ is naturally coherent with respect to sufficiently low-frequency signals, regardless of its connectivity.
\end{enumerate}
One additional feature of our analysis is that it can be further applied in settings where the composition of the network is unknown and only distributional information is present. More precisely, we, in Section \ref{sec:dymC}, consider a network where node dynamics are given by random transfer functions. As the network size grows, the coherent dynamics $\bar{g}(s)$, the harmonic mean of all node dynamics, converges in probability to a deterministic transfer function. We term such a phenomenon, where a family of uncertain large-scale systems concentrates to a common deterministic system, \emph{dynamics concentration}.

Lastly, we verify our theoretical results in Section \ref{sec:exmples} by several numerical experiments on linearized power network model, and discuss a general aggregation model for a group of coherent generators.

\section{Coherence in Frequency Domain}\label{sec:ptw_conv}
    
In this section, we analyze the network coherence as the low-rank structure of the transfer matrix in the frequency domain. We start with an important lemma revealing how such coherence is related to the algebraic connectivity $\lambda_2(L)$ and the coupling dynamics $f(s)$. 

\begin{lem}\label{lem_reg_norm_bd}
    Let $T(s)$ and $\bar{g}(s)$ be defined as in \eqref{eq_T_explict} and \eqref{eq_g_bar}, respectively. Suppose that for $s_0\in\compl$ that is not a pole of $f(s)$, we have $$|\bar{g}(s_0)|\leq M_1, \text{and }\max_{1\leq i\leq n}|g_i^{-1}(s_0)|\leq M_2\,,$$ for some $M_1,M_2>0$. Then the following inequality holds:
    \be
        \lV T(s_0)-\frac{1}{n}\bar{g}(s_0)\one\one^\top \rV\leq \frac{\lp M_1M_2+1\rp^2}{|f(s_0)|\lambda_2(L)-M_2-M_1M_2^2}\,,\label{eq_T_norm_bd}
    \ee
    whenever $|f(s_0)|\lambda_2(L)\geq M_2+M_1M_2^2$.
\end{lem}
We refer readers to Appendix \ref{app_pf_lem_reg_norm_bd} for the proof. Lemma 4 provides a non-asymptotic rate for our incoherence measure
\begin{equation}\label{eq:non-asymp-rate}
    \lV T(s_0)-\frac{1}{n}\bar{g}(s_0)\one\one^\top \rV\sim \mathcal{O}\lp \frac{M_1^2M_2^2}{|f(s_0)|\lambda_2(L)}\rp\,.
\end{equation}
A large value of $|f(s_0)|\lambda_2(L)$ is sufficient to have the incoherence measure small, and we term this quantity as \emph{effective algebraic connectivity}. We see that there are two possible ways to achieve such point-wise coherence: Either we increase the network algebraic connectivity $\lambda_2(L)$, by adding edges to the network and increasing edge weights, etc., or we move our point of interest $s_0$ to a pole of $f(s)$. This point-wise coherence via effective connectivity provides the basis of our subsequent analysis. 
As we mentioned above, we can achieve such coherence by increasing either $\lambda_2(L)$ or $|f(s_0)|$, provided that the other value is fixed and non-zero. Section \ref{ssec_coherence_high_con} considers the former and  Section \ref{ssec_coherence_high_gain} the latter.

\subsection{Coherence under high network connectivity}\label{ssec_coherence_high_con}
It is intuitive that a network behaves coherently under high connectivity. A formal frequency domain characterization is stated as follow.
\begin{thm}\label{thm_unifm_conv_reg_compact}
    Let $T(s)$ and $\bar{g}(s)$ be defined as in \eqref{eq_T_explict} and \eqref{eq_g_bar}, respectively. Given a compact set $S\subset \compl$, if
    \begin{enumerate}
        \item $S$ does not contain any zero or pole of $\bar{g}(s)$;
        \item $\inf_{s\in S}|f(s)|>0$\,,
    \end{enumerate}  
    we have $
        \lim_{\lambda_2(L)\ra +\infty}\sup_{s\in S}\lV T(s)-\frac{1}{n}\bar{g}(s)\one\one^\top \rV=0\,.$
\end{thm}
\begin{pf}
    On the one hand, since $S$ does not contain any pole of $\bar{g}(s)$, $\bar{g}(s)$ is continuous on the compact set $S$, and hence bounded~\cite[Theorem 4.15]{Rudin1964}. On the other hand, because $S$ does not contain any zero of $\bar{g}(s)$, every $g_i^{-1}(s)$ must be continuous on $S$, and hence bounded as well. It follows that $\max_{1\leq i\leq n}|g_i^{-1}(s)|$ is bounded on $S$, and the conditions of Lemma \ref{lem_reg_norm_bd} are satisfied for all $s\in S$ with a uniform choice of $M_1$ and $M_2$. By \eqref{eq_T_norm_bd}, we have
    \ben
        \sup_{s\in S}\lV T(s)-\frac{1}{n}\bar{g}(s)\one\one^\top \rV\leq 
        \frac{\lp M_1M_2+1\rp^2}{F_l\lambda_2(L)-M_2-M_1M_2^2}\,,
    \een
    where $F_{l}=\inf_{s\in S}|f(s)|$.
    We finish the proof by taking $\lambda_2(L)\ra +\infty$ on both sides.   
\end{pf}
Theorem \ref{thm_unifm_conv_reg_compact} formally shows that high network connectivity leads to coherence. We emphasize that such coherence is frequency-dependent: the incoherence measure is defined over a compact set $S$. Roughly speaking, if we would like to see whether the network could have coherent response under certain input signal, then $S$ should cover most of the frequency components of that signal, as well satisfies the assumptions in Theorem \ref{thm_unifm_conv_reg_compact}. We discuss the proper choice of $S$ when we use Theorem \ref{thm_unifm_conv_reg_compact} to infer the time-domain response in Section \ref{secc_to_time_high_con}.

\subsection{Coherence under high gain in coupling dynamics}\label{ssec_coherence_high_gain}
However, high network connectivity is not necessary for coherence. A high gain in the coupling dynamics effectively amplifies the network connection, leading to the following frequency-domain coherence. 
\begin{thm}\label{thm_ptw_singular_f_pole}
    Let $T(s)$ and $\bar{g}(s)$ be defined as in \eqref{eq_T_explict} and \eqref{eq_g_bar}, respectively. Given a pole of $f(s)$, if
    \begin{enumerate}
        \item $s_0$ is neither a pole nor a zero of $\bar{g}(s)$;
        \item $\lambda_2(L)>0$,
    \end{enumerate}
    then
    $
        \lim_{s\ra s_0} \lV T(s)-\frac{1}{n}\bar{g}(s)\one\one^\top \rV=0\,.
    $
\end{thm}
\begin{pf}
    Since $s_0$ is neither a zero nor a pole of $\bar{g}(s)$, $\exists \delta_1>0$ such that $\forall s\in \mathcal{B}(s_0,\delta_1)$, we have $|\bar{g}^{-1}(s)|\leq M_1$ and $\max_{1\leq i\leq n}|g_i^{-1}(s)|\leq M_2$ for some $M_1,M_2>0$.
    
    Now notice that $\lim_{s\ra s_0}|f(s)|=+\infty$, by the definition of the limit, we know that $\exists \delta_2>0$ such that $\forall s\in \mathcal{B}(s_0,\delta_2)$, we have 
    $\frac{1}{2}|f(s)|\lambda_2(L)\geq M_2+M_1M_2^2\,.$
    By Lemma \ref{lem_reg_norm_bd}, let $\delta:=\min\{\delta_1,\delta_2\}$, then $\forall s\in \mathcal{B}(s_0,\delta)$, the following holds
    \begin{align*}
        \lV T(s)-\frac{1}{n}\bar{g}(s)\one\one^\top \rV&\leq\; 
        \frac{\lp M_1M_2+1\rp^2}{|f(s)|\lambda_2(L)-M_2-M_1M_2^2}\\
        &\leq \; \frac{2\lp M_1M_2+1\rp^2}{|f(s)|\lambda_2(L)}\,.
    \end{align*}
    Taking $s\ra s_0$, the limit of right-hand side is 0. 
\end{pf}
Theorem \ref{thm_ptw_singular_f_pole} suggests that for any connected network, some coupling dynamics causes coherent responses from the network under specific input signals. For example, when $f(s)=\frac{1}{s}$, the network $T(s)$ is naturally coherent around $s=0$, which implies that such network behaves coherently under sufficiently low-frequency input signals. This is formally justified in Section \ref{ssec_to_time_high_gain}, along with time-domain results for other choice of coupling dynamics.
\begin{rem}
    The convergence results presented in this section exclude the region that contains any zero or pole of $\bar{g}(s)$. One can derive convergence results over those regions under certain conditions, but the results is less useful in understanding the network's time-domain behavior. \ifthenelse{\boolean{archive}}{We discuss these additional convergence results in Appendix \ref{app_add_conv}}{We refer readers to the technical note~\cite{min2021a} for details.}
\end{rem}
\section{Implications on Time-Domain Response}\label{sec_to_time}
In this section, we discuss how one can infer the network's time-domain response using the established frequency-domain coherence in Theorem \ref{thm_unifm_conv_reg_compact} and \ref{thm_ptw_singular_f_pole}. Provided that the network $T(s)$ and the coherent dynamics $\bar{g}(s)$ are BIBO stable, we let $\mathbf{y}(t)=[y_1(t),\cdots,y_i(t),\cdots,y_n(t)]^\top $ be the response of the network when the network input is $U(s)$, and let $\bar{y}(t)$ be the response of $\bar{g}(s)$ to $\frac{\one^\top }{n}U(s)$. The inverse Laplace transform~\cite{Dullerud2013} suggests that for all $i=1,\cdots,n$, we have
\begin{align}
    &\;|y_i(t)-\bar{y}(t)|=\nonumber\\ &\;\lvt\lim_{\omega\ra\infty}\int_{\sigma-j\omega}^{\sigma+j\omega}e^{st}e_i^\top \lp T(s)-\frac{1}{n}\bar{g}(s)\one\one^\top \rp U(s)ds\rvt\label{eq_inverse_lap}\,,
\end{align}
with a proper choice of $\sigma>0$. Here $e_i$ is the $i$-th column of identity matrix $I_n$. This integral can be decomposed in two parts: one integral on the low-frequency band $(\sigma-j\omega_0,\sigma+j\omega_0)$; and another on the high-frequency band $(\sigma-j\infty,\sigma-j\omega_0)\cup (\sigma+j\omega_0,\sigma+j\infty)$, with some choice of $\omega_0$. The former can be made small in absolute value by controlling the incoherence measure $\|T(s)-\frac{1}{n}\bar{g}(s)\one\one^\top \|$ over the set $S:(\sigma-j\omega_0,\sigma+j\omega_0)$. In particular,
\begin{enumerate}
    \item $\sup_{s\in S}\|T(s)-\frac{1}{n}\bar{g}(s)\one\one^\top \|$ can be small under high network connectivity, as suggested by Theorem \ref{thm_unifm_conv_reg_compact};
    \item $\sup_{s\in S}\|T(s)-\frac{1}{n}\bar{g}(s)\one\one^\top \|$ can be small when $S$ is confined in a neighborhood around pole of coupling dynamics $f(s)$, suggested by Theorem \ref{thm_ptw_singular_f_pole}. The case $f(s)=\frac{1}{s}$ is of the most interest.
\end{enumerate}
Moreover, when $U(s)$ is a sufficiently low-frequency signal such that the high-frequency band $(\sigma-j\infty,\sigma-j\omega_0)\cup (\sigma+j\omega_0,\sigma+j\infty)$ does not include much of its frequency components, the latter integral can be made small. Given an upper bound on the integral in \eqref{eq_inverse_lap}, we show that the time-domain response of every node in the network resembles the one from the coherent dynamics $\bar{g}(s)$. Similar to Section \ref{sec:ptw_conv}, we show such time-domain coherence in two regimes: high network connectivity or high gain in the coupling dynamics.
\begin{rem}
    In order to infer the time-domain response, it is necessary that both the transfer functions $T(s)$ and $\frac{1}{n}\bar g(s)\one\one^\top $ are stable. Since our primary focus is on the interpretation of the frequency domain results, we are largely working under the tacit assumption that these transfer functions are stable whenever required. 
    It should also be noted that there exist a range of scalable stability criteria in the literature that can be used to guarantee internal stability of the feedback setup in Figure~\ref{blk_p_n}. Perhaps the most well known is that if each $g_i(s)$ is \textit{strictly positive real}, and $f(s)$ is \textit{positive real}, then the transfer functions $\bar{g}(s)$ and
    \[
    \begin{bmatrix}
    G(s)\\I
    \end{bmatrix}\left(I+f(s)LG(s)\right)^{-1}\begin{bmatrix}
    f(s)L&I
    \end{bmatrix}
    \]
    are stable (see e.g. \cite{MD95}). Alternative approaches that can be easily adapted to our framework that give criteria that allow for different classes of transfer functions include \cite{LV06,JK10,pm19tcns}.
\end{rem}
\subsection{Coherent response under high network connectivity}\label{secc_to_time_high_con}
Our first result considers network with high connectivity.
\begin{thm}\label{thm_to_time_high_con}
    Given a network with node dynamics $\{g_i(s)\}_{i=1}^n$ and coupling dynamics $f(s)$, assume that there exists $\gamma>0$, such that $\|\bar{g}(s)\|_{\mathcal{H}_\infty}\leq \gamma$ and $\|T(s)\|_{\mathcal{H}_\infty}\leq \gamma$ for any symmetric Laplacian matrix $L$. Consider a network coupling $f(s)$ and a real input signal vector $\mathbf{u}(t)$ with its Laplace transform $U(s)$ such that for some $\sigma>0$, we have
    \begin{enumerate}
        \item $\inf_{\omega\in\mathbb{R}}|f(\sigma+i\omega)|>0$;
        \item $\sup_{Re(s)>\sigma}\|U(s)\|$ is finite;
        \item $
         \lim_{\omega\ra \infty}\int_{\sigma+j0}^{\sigma+j\omega} \|U(s)\|ds
        $ is finite\,.
    \end{enumerate}
    Then for any $\epsilon>0$, there exists a $\lambda>0$, such that whenever $\lambda_2(L)\geq \lambda$, we have $\|\mathbf{y}(t)-\bar{y}(t)\one\|_{\mathcal{L_\infty}}\leq \epsilon$, i.e.,
    $$
        \max_{i\in[n]}\sup_{t>0}|y_i(t)-\bar{y}(t)|\leq \epsilon\,.
    $$
\end{thm}
We refer readers to Appendix \ref{thm_to_time_pf} for the proof.  Theorem \ref{thm_to_time_high_con} provide a formal explanation of coherent behavior observed in practical networks and show its relation with network connectivity. That is, a stable network with high connectivity can respond coherently to a class of input signals. More importantly, the coherently response is well approximated by $\bar{g}(s)$, then it suffices to study $\bar{g}(s)$ for understanding the coherent behavior of a network with high connectivity.

While the theorem suggests that some level of coherence can be achieved by increasing the network connectivity, one should be cautious about the potential network instability caused by strong interconnection. Nonetheless, some simple passivity motivated criteria that ensure stability even as $\lambda_2(L)$ becomes arbitrarily large:

\begin{thm}\label{thm_passive_to_stable}
    Suppose that all $g_i(s),i=1,\cdots,n$ are \emph{output strictly passive}: $
        Re(g_i(s))\geq \epsilon |g_i(s)|^2, \forall Re(s)>0\,,
    $ for some $\epsilon>0$, and $f(s)$ is \emph{positive real}: $
        Re(f(s))\geq 0, \forall Re(s)>0\,,
    $
    then there exists $\gamma>0$, such that given any positive semidefinite matrix $L$, we have
    $$
        \|\bar{g}(s)\|_{\mathcal{H}_\infty}\leq \gamma,\ \mathrm{and}\ \|T(s)\|_{\mathcal{H}_\infty}\leq \gamma\,.
    $$
\end{thm}
We refer readers to Appendix \ref{thm_to_passive_to_stable_pf} for the proof. Theorem \ref{thm_passive_to_stable}, together with Theorem \ref{thm_to_time_high_con}, shows that for certain passive networks, the coherence can be achieved over a class of input signals by increasing the network connectivity.
\begin{rem}
    Besides network stability as a prerequisite, a few assumptions are made: infimum on $|f(s)|$ ensures that the network coupling does not vanish over our domain of interest; supremum on $\|U(s)\|$ is needed for utilizing inverse Laplace transform; and the last assumption requires $U(s)$ to have light tail on the high-frequency range, a low-frequency signal with no abrupt change at $t=0$, such as sinusoidal signal $U(s)=\frac{\alpha}{s^2+\alpha^2}\mathbf{u}_0$, or exponential approach signal $U(s)=\frac{\alpha}{s(s+\alpha)}\mathbf{u}_0$ of some shape $\mathbf{u}_0\in\mathbb{R}^n$, satisfies the assumption.
\end{rem}
\subsection{Coherent response under special coupling dynamics}\label{ssec_to_time_high_gain}
As we discussed in Section \ref{sec:ptw_conv}, coherence is not all about network connectivity, and high gain in the coupling dynamics causes coherence as well. One simple and practically seen coupling dynamics are $f(s)=\frac{1}{s}$. Due to its high gain at $s=0$, we expected the a coherent response under low-frequency signals, as formally shown below.
\begin{thm}\label{thm_to_time_high_gain}
    Given a network with node dynamics $\{g_i(s)\}_{i=1}^n$, coupling dynamics $f(s)=\frac{1}{s}$, and a fixed graph Laplacian $L$ with $\lambda_2(L)>0$, such that $\|\bar{g}(s)\|_{\mathcal{H}_\infty}$ and $\|T(s)\|_{\mathcal{H}_\infty}$ are finite, 
    we let the network input be a sinusoidal signal $\mathbf{u}_\alpha(t)=\sin(\alpha t)\chi(t)\mathbf{u}_0$ in an arbitrary direction $\mathbf{u}_0\in\mathbb{S}^{n-1}$. 
    Then for any $\epsilon>0$, there exists an $\alpha_0>0$ such that whenever $0\leq\alpha\leq \alpha_0$, we have $\|\mathbf{y}(t)-\bar{y}(t)\one\|_{\mathcal{L_\infty}}\leq \epsilon$, i.e.,
    \be
        \max_{i\in[n]}\sup_{t>0}|y_i(t)-\bar{y}(t)|\leq \epsilon\,.
    \ee
\end{thm}
We refer readers to Appendix \ref{thm_to_time_pf} for the proof. Theorem \ref{thm_to_time_high_gain} shows that a stable network with $f(s)=\frac{1}{s}$ is naturally coherent subject to sufficiently low-frequency signals, regardless of its connectivity. Notably, the requirement on the node dynamics here is much weaker than one in Theorem \ref{thm_to_time_high_con} as we only need to establish stability for a given interconnection $L$, whereas Theorem \ref{thm_to_time_high_con} requires stability under any interconnection.

\subsection{Comparison with different notions of coordination}\label{ssec_comp_time}
Our Theorem \ref{thm_to_time_high_con} and \ref{thm_to_time_high_gain} shows the coherent response of network in time domain. We compare our results to prior work that studies different forms of time-domain coordination in network systems.

The consensus~\cite{Olfati-Saber20041520} and synchronization~\cite{Mirollo1990,mmhzt2015ton,Sepulchre2008706} is arguably the simplest form of coordination in network systems, which can be viewed as a problem tracking some reference signal $\bar{y}(t)$ representing the final consensus or synchronization. However, one only requires $y_i(t)\ra \bar{y}(t)$ when $t\ra \infty$, i.e., that the node responses become close to $\bar{y}(t)$ in steady state. The coherent response considered here is different in that we have $y_i(t)\simeq \bar{y}(t), \forall t>0$, i.e., $\bar{y}(t)$ is a good approximation for $y_i(t)$ for all time $t>0$, hence our results can be also used for transient analysis. 

The work on coherency and synchrony~\cite{Ramaswamy1995,ramaswamy1996,Wu1983,Sastry1981} study a similar behavior as us, but characterized as pairwise coherence achieved under input signal of certain spatial shape: given a input signal vector $\mathbf{u}(t)=v(t)\mathbf{u}_0$,~\cite{Ramaswamy1995,Wu1983} shows the condition on $\mathbf{u}_0$ such that the responses of some pair of nodes are similar (or generally, proportional~\cite{ramaswamy1996}), i.e., $y_i(t)\simeq y_j(t)$ for some $i,j\in[n]$ . Our results show that certain temporal shape $v(t)$ also causes coherence, and in a stronger form: our coherence does not depends on the shape $u_0$, and holds for all nodes. 



\section{Dynamics Concentration in Large-scale Networks}\label{sec:dymC}
    
    In Section \ref{sec:ptw_conv}, we looked into convergence results of $T(s)$ for networks with fixed size $n$. However, one could easily see that such coherence depends mildly on the network size $n$: In Lemma \ref{lem_reg_norm_bd}, as long as the bounds regarding $g_i(s)$, i.e. $M_1$ and $M_2$ do not scale with respect to $n$, coherence can emerge as the network size increases. This is the topic of this section.
    
    \subsection{Coherence in large-scale networks}
    To start with, we revise the problem settings to account for variable network size: Let $\{g_i(s), i\in \mathbb{N}_+\}$ be a sequence of transfer functions, and $\{L_n, n\in\mathbb{N}_+\}$ be a sequence of real symmetric Laplacian matrices such that $L_n$ is a square matrix of order $n$, particularly, let $L_1=0$. Then we define a sequence of transfer matrix $T_n(s)$ as
    \be
        T_n(s) = \lp I_n+G_n(s)L_n\rp^{-1} G_n(s)\,,\label{eq_T_n_explicit}
    \ee
    where $G_n(s)= \dg\{g_1(s),\cdots,g_n(s)\}$. This is exactly the same transfer matrix  shown in Fig.\ref{blk_p_n} for a  network of size $n$. We can then define the coherent dynamics for every $T_n(s)$ as
    $
        \bar{g}_n(s) = \lp\frac{1}{n}\sum_{i=1}^ng_i^{-1}(s)\rp^{-1}
    $.

    For certain family $\{L_n,n\in\mathbb{N}_+\}$ of large-scale networks, the network algebraic connectivity $\lambda_2(L_n)$ increases as $n$ grows. For example, when $L_n$ is the Laplacian of a complete graph of size $n$ with all edge weights being $1$, we have $\lambda_2(L_n)=n$. As a result, network coherence naturally emerges as the network size grows. Recall that to prove the convergence of $T_n(s)$ to $\frac{1}{n}\bar{g}_n(s)\one\one^\top $ for fixed $n$, we essentially seek for $M_1,M_2>0$, such that $|\bar{g}_n(s)|\leq M_1$ and $\max_{1\leq i\leq n}|g_i^{-1}(s)|\leq M_2$ for $s$ in a certain set. If it is possible to find a universal $M_1,M_2>0$ for all $n$, then the convergence results should be extended to arbitrarily large networks, provided that network connectivity increases as $n$ grows. The results follows after we state the notion of uniform boundedness for a family of functions.
    \begin{defn}
    Let $\{g_i(s), i\in I\}$ be a family of complex functions indexed by $I$. Given $S\subset \compl$, $\{g_i(s), i\in I\}$ is uniformly bounded on $S$ if
    \ben
        \exists M>0\quad s.t.\quad |g_i(s)|\leq M,\quad \forall i\in I,\ \forall s\in S\,. 
    \een
    \end{defn}
    \begin{thm}\label{thm_unifm_conv_reg_dymC}
    Suppose $\lambda_2(L_n)\ra +\infty$ as $n\ra \infty$. Given a compact set $S\subset\mathbb{C}$, if both $\{g_i^{-1}(s),i\in\mathbb{N}_+\}$ and $\{\bar{g}_n(s),n\in\mathbb{N}_+\}$ are uniformly bounded on a set $S\subset \compl$, and $inf_{s\in S}|f(s)|>0$, then we have
    \ben
        \lim_{n\ra \infty}\sup_{s\in S}\lV T_n(s)-\frac{1}{n}\bar{g}_n(s)\one\one^\top \rV=0\,.
    \een
    \end{thm}
    \ifthenelse{\boolean{archive}}{
    \begin{pf}
    Since both $\{g_i^{-1}(s),i\in\mathbb{N}_+\}$ and $\{\bar{g}_n(s),n\in\mathbb{N}_+\}$ are uniformly bounded on $S$, $\exists M_1,M_2>0$ s.t. $|\bar{g}_n(s)|\leq M_1$ and $\max_{1\leq i\leq n}|g_i^{-1}(s)|\leq M_2$ for every $n\in \mathbb{N}_+$ and $s\in S$. By Lemma \ref{lem_reg_norm_bd}, $\forall n\in\mathbb{N}_+$,
    \be
        \sup_{s\in S}\lV T_n(s)-\frac{1}{n}\bar{g}_n(s)\one\one^\top \rV\leq 
        \frac{\lp M_1M_2+1\rp^2}{F_l\lambda_2(L_n)-M_2-M_1M_2^2}\,,\label{eq_T_norm_bd_dymC}
    \ee
    where $F_l=\inf_{s\in S}|f(s)|$. We already assumed that $\lambda_2(L_n)\ra +\infty$ as $n\ra +\infty$, therefore the proof is finished by taking $n\ra +\infty$ on both sides of \eqref{eq_T_norm_bd_dymC}.
    \end{pf}
    }{The proof is similar to the one for Theorem \ref{thm_unifm_conv_reg_compact}. Due to the space constraints, we refer readers to the technical note~\cite{min2021a} for the proof.} Interestingly, in a stochastic setting where all $g_i(s)$ are unknown transfer functions independently drawn from some distribution, their harmonic mean $\bar{g}_n(s)$ eventually converges in probability to a deterministic transfer function as the network size increases. Consequently, a large-scale network consisting of random node dynamics (to be formally defined later) concentrates to deterministic a system. We term this phenomenon \emph{dynamics concentration}.
    \begin{rem}
        In this section, we only discuss the coherence due to connectivity, since the coherence from high gain in coupling dynamics shown in Theorem \ref{thm_ptw_singular_f_pole} can be applied to any connected network, regardless of its size.
    \end{rem}
    \subsection{Dynamics concentration in large-scale networks}
    Now we consider the cases where the node dynamics are unknown (stochastic). For simplicity, we constraint our analysis to the setting where the node dynamics are independently sampled from the same random rational transfer function with all or part of the coefficients are random variables, i.e. the nodal transfer functions are of the form
    \begin{equation}\label{eq_random_tf}
        g_i(s) \sim \frac{b_ms^m+\dots b_1s+b_0}{a_ls^l+\dots a_1s+a_0}\,,
    \end{equation}
    for some $m,l>0$, where $b_0,\cdots,b_m$, $a_0,\cdots, a_l$ are random variables. 

    To formalize the setting, we firstly define the random transfer function to be sampled. Let $\Omega=\mathbb{R}^d$ be the sample space, $\mathcal{F}$ the Borel $\sigma$-field of $\Omega$, and $\mathbb{P}$ a probability measure on $\Omega$. A sample $w\in\Omega$ thus represents a $d$-dimensional vector of coefficients. We then define a random rational transfer function $g(s,w)$ on $(\Omega,\mathcal{F},\mathbb{P})$ such that all or part of the coefficients of $g(s,w)$ are random variables. Then for any $w_0\in \Omega$, $g(s,w_0)$ is a rational transfer function.
    
    Now consider the probability space $(\Omega^\infty,\mathcal{F}^\infty,\mathbb{P}^\infty)$. Every $\mathbf{w}\in \Omega^\infty$ give an instance of samples drawn from our random transfer function: $$g_i(s,w_i):=g(s,w_i),i\in\mathbb{N}_+\,,$$ where $w_i$ is the $i$-th element of $\mathbf{w}$. By construction, $g_i(s,w_i), i\in\mathbb{N}_+$ are i.i.d. random transfer functions. Moreover, for every $s_0\in\compl$, $g_i(s_0,w_i),i\in\mathbb{N}_+$ are i.i.d. random complex variables taking values in the extended complex plane (presumably taking value $\infty$).

    Now given $\{L_n, n\in\mathbb{N}_+\}$ a sequence of $n\by n$ real symmetric Laplacian matrices, consider the random network of size $n$ whose nodes are associated with the dynamics $g_i(s,w_i),i=1,2,\cdots,n$ and coupled through $L_n$. The transfer matrix of such a network is given by
    \be\label{eq_T_stoch}
        T_n(s,\mathbf{w})=(I_n+G_n(s,\mathbf{w})L_n)^{-1}G_n(s,\mathbf{w})\,,
    \ee
    where $G_n(s,\mathbf{w})=\dg\{g_1(s,w_1),\cdots,g_n(s,w_n)\}$. Then under this setting, the coherent dynamics of the network is given by
    \be\label{eq_g_bar_stoch}
        \bar{g}(s,\mathbf{w}) = \lp \frac{1}{n}\sum_{i=1}^ng_i^{-1}(s,w_i)\rp^{-1}\,.
    \ee
    Now given a compact set $S\subset \compl$ of interest, and assuming suitable conditions on the distribution of $g(s,w)$, we expect that the random coherent dynamics $\bar{g}(s,\mathbf{w})$ would converge uniformly in probability to its expectation
    \be\label{eq_expc_g}
        \hat{g}(s)=\lp\expc g^{-1}(s,w))\rp^{-1}:=\lp \int_{\Omega}g^{-1}(s,w)d\mathbb{P}(w)\rp^{-1}\,,
    \ee
    for all $s\in S$, as $n\ra \infty$. The following Lemma provides a sufficient condition for this to hold. 
    \begin{lem}\label{lem_unifm_prob_conv_compact}
    Consider the probability space $(\Omega^\infty,\mathcal{F}^\infty,\mathbb{P}^\infty)$. Let $\bar{g}_n(s,\mathbf{w})$ and $\hat{g}(s)$ be defined as in \eqref{eq_g_bar_stoch} and \eqref{eq_expc_g}, respectively, and given a compact set $S\subset \compl$, let the following conditions hold:
    \begin{enumerate}
        \item $g^{-1}(s,w)$ is uniformly bounded on $S\times \Omega$;
        \item $\{\bar{g}_n(s,\mathbf{w}),n\in\mathbb{N}_+\}$ are uniformly bounded on $S\times \Omega^\infty$;
        \item $\exists L>0$ s.t. $|g_1^{-1}(s_1,w)-g_1^{-1}(s_2,w)|\leq L|s_1-s_2|$, $\forall w\in\Omega,\forall s_1,s_2\in S$;
        \item $\hat{g}(s)$ is uniformly continuous.
    \end{enumerate}
    Then, $\forall \epsilon>0$, we have
    \ben
        \lim_{n\ra \infty}\mathbb{P}\lp\sup_{s\in S}\lV \frac{1}{n}\bar{g}_n(s,\mathbf{w})\one\one^\top -\frac{1}{n}\hat{g}(s)\one\one^\top \rV\geq \epsilon\rp=0\,.
    \een
\end{lem}
    This lemma suggests that our coherent dynamics $\bar{g}_n(s,\mathbf{w})$, as $n$ increases, converges uniformly on $S$ to its expected version $\hat{g}(s)$. Then provided that the coherence is obtained as the network size grows, we would expect that the random transfer matrix $T_n(s, \mathbf{w})$ to concentrate to a deterministic one $\frac{1}{n}\hat{g}(s)\one\one^\top $, as the following theorem shows.
    \begin{thm}\label{thm_unifm_prob_conv_compact}
    Given probability space $(\Omega^\infty,\mathcal{F}^\infty,\mathbb{P}^\infty)$. Let $T_n(s,\mathbf{w})$ and $\hat{g}(s)$ be defined as in \eqref{eq_T_stoch} and \eqref{eq_expc_g}, respectively. Suppose $\lambda_2(L_n)\ra +\infty$ as $n\ra +\infty$. Given a compact set $S\subset \compl$, if all the conditions in Lemma \ref{lem_unifm_prob_conv_compact} hold, then $\forall \epsilon>0$, we have
    \ben
        \lim_{n\ra \infty}\mathbb{P}\lp\sup_{s\in S}\lV T_n(s,\mathbf{w})-\frac{1}{n}\hat{g}(s)\one\one^\top \rV\geq \epsilon\rp=0\,.
    \een
\end{thm}
    \ifthenelse{\boolean{archive}}{The proofs of Lemma \ref{lem_unifm_prob_conv_compact}  and Theorem \ref{thm_unifm_prob_conv_compact} are shown in  Appendix \ref{app_proof_lem_unifm_prob_conv_compact}.}{The proof of Lemma \ref{lem_unifm_prob_conv_compact} follows the standard procedure for showing the uniform stochastic convergence of a random function, then Theorem \ref{thm_unifm_prob_conv_compact} is its direct application. We refer interested readers to the technical note~\cite{min2021a} for the proofs.}
    In summary, because the coherent dynamics is given by the harmonic mean of all node dynamics $g_i(s)$, it concentrates to its harmonic expectation $\hat g(s)$ as the network size grows. As a result, in practice, the coherent behavior of large-scale networks depends on the empirical distribution of $g_i(s)$, i.e. a collective effect of all node dynamics rather than every individual node dynamics. For example, two different realizations of large-scale network with dynamics $T_n(s,\mathbf{w})$ exhibit similar coherent behavior with high probability, in spite of the possible substantial differences in individual node dynamics.       
    \begin{rem}
        With Theorem \ref{thm_unifm_prob_conv_compact}, one can adopt the analysis in Section \ref{sec_to_time} to derive a time-domain result similar to the one in Theorem \ref{thm_to_time_high_con}. In this case, the network stability again relies on node passivity as required in Theorem \ref{thm_passive_to_stable}. Nonetheless, for low-order rational transfer function, the condition of being passive is equivalent to its coefficients satisfying certain algebraic inequalities\cite{chen2009}, hence there exists probability measure $\mathbb{P}$ on the coefficients such that the resulting transfer function is passive almost surely, under which the time-domain response of the network $T_n(s,\mathbf{w})$ can be inferred.  
    \end{rem}
\section{Application: Aggregate Dynamics of Synchronous Generator Networks}\label{sec:exmples}
In this section, 
we apply our analysis to investigate coherence in power networks. For coherent generator groups, we find that $\frac{1}{n}\bar{g}(s)$ generalizes typical aggregate generator models which are often used for model reduction in power networks~\cite{Chow2013}. Moreover, we show that heterogeneity in generator dynamics usually leads to high-order aggregate dynamics, making it challenging to find a reasonably low-order approximation. 
Consider the transfer matrix of power generator networks~\cite{Paganini2019tac} linearized around its steady-state point, given by the following block diagram:
\begin{figure}[ht]
    \centering
    \includegraphics[height=2.5cm]{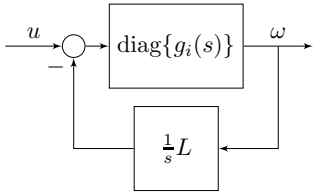}
    \caption{Block Diagram of Linearized Power Networks}\label{blk_power}
\end{figure}
This is exactly the block structure shown in Fig. \ref{blk_p_n} with $f(s)=\frac{1}{s}$. Here, the network output, i.e., the frequency deviation of each generator, is denoted by $\omega$. Generally, the $g_i(s)$ are modeled as strictly positive real transfer functions and we assume $L$ is connected. Such interconnection is stable~\cite{MD95}, regardless of the network connectivity. 
\begin{figure*}[!t]
  \includegraphics[width=0.95\textwidth]{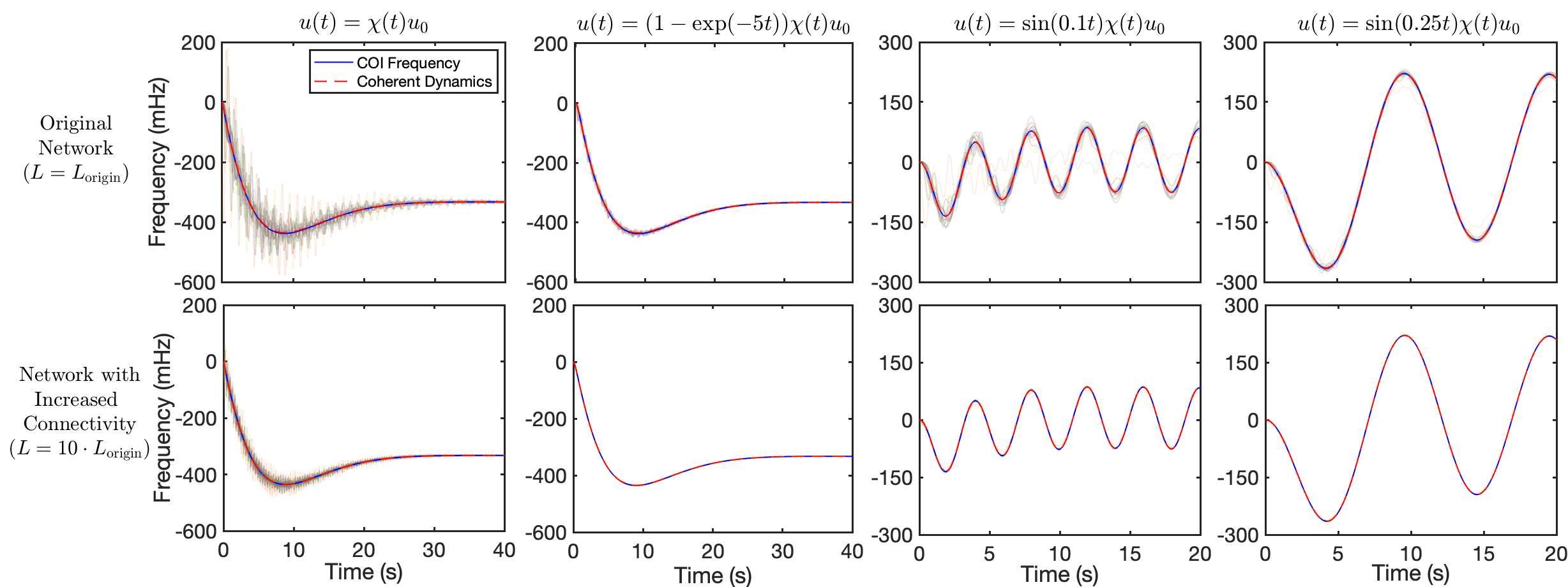}
  \caption{Coherent response of Icelandic Grid. Each column corresponds to a different input signal (from left to right: step, exponential approach, high-frequency sinusoidal, and low-frequency sinusoidal signal); The input signal has a shape $u_0=-e_2$, i.e., only the second node is subject to disturbance. Top row shows the responses of original icelandic grid, and the bottom row shows the responses of network with increased connectivity. Red dashed line shows the response of $\bar{g}(s)$ subject to the averaged input $\bar{u}(t)=\one^\top u(t)/n$. Blue solid line shows the Center-of-Inertia frequency of the grid $y_{\mathrm{COI}}=(\sum_{i=1}^nm_iy_i)/(\sum_{i=1}^nm_i)$.}
  \label{fig_coherence}
\end{figure*}    
\subsection{Numerical verification}
We verify our theoretical results, Theorem \ref{thm_to_time_high_con} and Theorem \ref{thm_to_time_high_gain}, with numerical simulations on the Icelandic power grid~\cite{iceland} modeled as in Fig \ref{blk_power}. We plot  in Fig.~\ref{fig_coherence} the frequency response of the power network model subject to various input disturbances. the network step response is more coherent, i.e. response of every single node (generator) is getting closer to the one of the coherent dynamics $\bar{g}(s)$, when the network connectivity is scaled up, as suggested by Theorem \ref{thm_to_time_high_con}. In addition, the network responds more coherently when subject to lower-frequency signals (See the second and forth column in Fig \ref{fig_coherence}), as suggested by Theorem \ref{thm_to_time_high_gain}. But most importantly, the coherent dynamics $\bar{g}(s)$ provides a good characterization of the coherent response. We also plot the Center-of-Inertia frequency of the grid $y_{\mathrm{COI}}=(\sum_{i=1}^nm_iy_i)/(\sum_{i=1}^nm_i)$, which is generally used for frequency response assessment, and we see that it is well approximated by the response of $\bar{g}(s)$.

\subsection{Aggregate dynamics of generator networks}    
The numerical simulations above suggest that the coherent dynamics $\bar g(s)$ characterize well the overall frequency response of generators in a grid. This leads to a general methodology to analyze the aggregate dynamics of such networks. Let $$g_\mathrm{aggr}(s):=\frac{1}{n}\bar{g}(s)=\lp \sum_{i=1}^ng_i^{-1}(s)\rp\,.$$ Our analysis suggests that the transfer function $T(s)$ representing a network of  generators is close $g_\mathrm{aggr}(s)\one\one^\top $ within the low-frequency range, for sufficiently high network connectivity $\lambda_2(L)$. We can also view $g_\mathrm{aggr}(s)$ as the aggregate generator dynamics, in the sense that it takes the sum of disturbances $\one^\top u=\sum_{i=1}^nu_i$ as its input, and its output represents the coherent response of all generators.
    
Such a notion of aggregate dynamics is important in modeling large-scale power networks\cite{Chow2013}. Generally speaking, one seeks to find an aggregate dynamic model for a group of generators using the same structure (transfer function) as individual generator dynamics, i.e. when generator dynamics are modeled as $g_i(s)=g(s;\theta_i)$, where $\theta_i$ is a vector of parameters representing physical properties of each generator, existing works~\cite{Germond1978,Guggilam2018} propose methods to find aggregate dynamics of the form $g(s;\theta_\mathrm{aggr})$ for certain structures of $g(s;\theta)$. Our $g_\mathrm{aggr}(s)$ justifies their choices of $\theta_\mathrm{aggr}$, as shown in the following example.
    
\begin{exmp}
    For generators given by the swing model $g_i(s) =\frac{1}{m_is+d_i}\,,$
    where $m_i,d_i$ are the inertia and damping of generator $i$, respectively. The aggregate dynamics are
    \be
        g_\mathrm{aggr}(s) = \frac{1}{m_\mathrm{aggr}s+d_\mathrm{aggr}}\,,\label{eq_aggr_dym_sw}
    \ee
    where $m_\mathrm{aggr} = \sum_{i=1}^nm_i$ and $d_\mathrm{aggr} = \sum_{i=1}^nd_i$.
\end{exmp}
Here the parameters are $\theta=\{m,d\}$. The aggregate model given by \eqref{eq_aggr_dym_sw} is consistent with the existing approach of choosing inertia $m$ and damping $d$ as the respective sums over all the coherent generators. 
    
However, as we show in the next example, when one considers more involved models,  it is challenging to find parameters that accurately fit the aggregate dynamics.
\begin{exmp}
    For generators given by the swing model with turbine droop
    $
        g_i(s) = \frac{1}{m_is+d_i+\frac{r_i^{-1}}{\tau_is+1}}\,,
    $
    where $r_i^{-1}$ and  $\tau_i$ are the droop coefficient and turbine time constant of generator $i$, respectively. The aggregate dynamics are given by 
    \be
        g_\mathrm{aggr}(s) = \frac{1}{m_\mathrm{aggr}s+d_\mathrm{aggr}+\sum_{i=1}^n\frac{r_i^{-1}}{\tau_is+1}}\,.\label{eq_aggr_dym_sw_tb}
    \ee
\end{exmp}
Here the parameters are $\theta=\{m,d,r^{-1},\tau\}$. This example illustrates, in particular, the difficulty in aggregating generators with heterogeneous turbine time constants. 
If the $\tau_i$  are heterogeneous, then $g_\mathrm{aggr}(s)$ is a high-order transfer function and cannot be accurately represented by a single generator model parametrized by $\theta$. The aggregation of generators essentially asks for a low-order approximation of $g_\mathrm{aggr}(s)$. Our analysis reveals the fundamental limitation of using conventional approaches seeking aggregate dynamics with the same structure of individual generators. Furthermore, by characterizing the aggregate dynamics in the explicit form $g_\mathrm{aggr}(s)$, one can develop more accurate low-order approximation~\cite{min2020lcss}. Lastly, we emphasize that our analysis does not depend on a specific model of generator dynamics $g_i(s)$, hence it provides a general methodology to aggregate coherent generator networks.

\section{Conclusions}\label{sec:conclusion}    
In this paper, we studies network coherence as a low-rank property of the transfer matrix $T(s)$ in the frequency domain. The analysis leads to useful characterizations of coordinated behavior and justifies the relation between network coherence and network effective algebraic connectivity. Our results suggest that network coherence is a frequency-dependent phenomenon, which is numerically illustrated in generator networks. Lastly, concentration results for large-scale networks are presented, revealing the exclusive role of the statistical distribution of node dynamics in determining the coherent dynamics of such networks. One interesting future work is to study the dynamic behavior of large-scale networks with multiple coherent groups. One could model the inter-community interactions by replacing the dynamics of each community with its coherent one, or more generally, a reduced one. Although clustering, i.e. finding communities, for homogeneous networks can be efficiently done by various graph-based methods, it is still open for research to find multiple coherent groups in heterogeneous dynamical networks.     

\appendix

\section{Proof of Lemma \ref{lem_reg_norm_bd}}\label{app_pf_lem_reg_norm_bd}
\begin{pf}
    Let $H=V^\top \dg\{g_i^{-1}(s_0)\}V+f(s_0)\Lambda$, such that 
    \eqref{eq_T_eigenform} becomes $
    T(s) = VH^{-1}V^\top $.
    Then it is easy to see that
    \begin{align}
        \lV T(s_0)-\frac{1}{n}\bar{g}(s_0)\one\one^\top \rV&=\; \|T(s_0)-\bar{g}(s_0)Ve_1e_1^\top V^\top \|\nonumber\\
        &=\; \lV V\lp H^{-1}-\bar{g}(s_0)e_1e_1^\top \rp V^\top \rV\nonumber\\
        &=\; \lV H^{-1}-\bar{g}(s_0)e_1e_1^\top  \rV\,,\label{eq_T_H_norm_equiv}
    \end{align}
    where $e_1$ is the first column of identity matrix $I_n$. The first equality holds by noticing that $\frac{\one}{\sqrt{n}}$ is the first column of $V$.
    
    With $V=\bmt
        \frac{\one}{\sqrt{n}}& V_\perp\emt$, we write $H$ in block matrix form:
    \begin{align*}
        H
        &= \bmt
            \frac{\one^\top }{\sqrt{n}}\\
            V_\perp^\top 
        \emt \dg\{g_i^{-1}(s_0)\} \bmt
        \frac{\one}{\sqrt{n}}& V_\perp\emt+f(s_0)\Lambda\\
        &:= \bmt
        \bar{g}^{-1}(s_0)& h^\top _{21}\\
        h_{21} & H_{22}
        \emt\,,
    \end{align*}
    where  
    \begin{align*}
        &\;h_{21}=V_\perp^\top \dg\{g_i^{-1}(s_0)\}\frac{\one}{\sqrt{n}}\,,\\
        &\;H_{22}=V_\perp^\top \dg\{g_i^{-1}(s_0)\}V_\perp+f(s_0)\Tilde{\Lambda}\,,\\
        &\;\Tilde{\Lambda}=\dg\{\lambda_2(L),\cdots,\lambda_n(L)\}\,.
    \end{align*}
    Inverting $H$ in its block form, we have
    \ben
        H^{-1} = \bmt
        a &-ah_{21}^\top H_{22}^{-1}\\
        -aH_{22}^{-1}h_{21}& H_{22}^{-1}+aH_{22}^{-1}h_{21}h_{21}^\top H_{22}^{-1}
        \emt\,,
    \een
    where $a = \frac{1}{\bar{g}^{-1}(s_0)-h_{21}^\top H_{22}^{-1}h_{21}}$.
    
    By our assumption, we have $\|\dg\{g_i^{-1}(s_0)\}\|=\max_{1\leq i\leq n}|g_i^{-1}(s_0)|\leq M_2\,,$ then
    \begin{align}
        \|h_{21}\|&=\; \lV V_\perp^\top \dg\{g_i^{-1}(s_0)\}\frac{\one}{\sqrt{n}}\rV\nonumber\\
        &\leq\; \|V_\perp\|\|\dg\{g_i^{-1}(s_0)\}\|\frac{\|\one\|}{\sqrt{n}}\leq M_2\,,\label{eq_h12_norm_bd}
    \end{align}
    and 
    \begin{align}
        \|H_{22}^{-1}\|&=\;\|(f(s_0)\Tilde{\Lambda}+ V_\perp^\top \dg\{g_i^{-1}(s_0)\}V_\perp)^{-1}\|\nonumber\\
        &=\;\sigma_{\min}\lp f(s_0)\Tilde{\Lambda}+ V_\perp^\top \dg\{g_i^{-1}(s_0)\}V_\perp\rp\nonumber\\
        &\leq\; \frac{1}{\sigma_{\min}(f(s_0)\Tilde{\Lambda})-\|V_\perp^\top \dg\{g_i^{-1}(s_0)\}V_\perp\|}\nonumber\\
        &\leq\; \frac{1}{\sigma_{\min}(f(s_0)\Tilde{\Lambda})-M_2}\leq \frac{1}{|f(s_0)|\lambda_2(L)-M_2} \,,\label{eq_H22_norm_bd}
    \end{align}
    whenever $|f(s_0)|\lambda_2(L)>M_2$.
    
    Lastly, when $|f(s_0)|\lambda_2(L)>M_2+M_2^2M_1$, a similar reasoning as above, using \eqref{eq_h12_norm_bd} \eqref{eq_H22_norm_bd}, and our assumption $|\bar{g}(s_0)|\leq M_1$, gives
    \begin{align}
        |a|
        &\leq\; \frac{1}{|\bar{g}^{-1}(s_0)|-\|h_{21}\|^2\|H_{22}^{-1}\|}\nonumber\\
        &= \;
        \frac{(|f(s_0)|\lambda_2(L)-M_2)M_1}{|f(s_0)|\lambda_2(L)-M_2-M_1M_2^2}\,.\label{eq_a_norm_bd}
    \end{align}
    Now we bound the norm of $H^{-1}-\bar{g}(s_0)e_1e_1^\top $ by the sum of norms of all its blocks:
    \begin{align}
        &\;\|H^{-1}-\bar{g}(s_0)e_1e_1^\top \|\nonumber\\
        \leq &\; |a\bar{g}(s_0)h_{21}^\top H_{22}^{-1}h_{21}|+2\|aH_{22}^{-1}h_{21}\|\nonumber\\
        &\; \quad\qquad+\|H_{22}^{-1}+aH_{22}^{-1}h_{21}h_{21}^\top H_{22}^{-1}\|\nonumber\\
        \leq &\; |a|\|H_{22}^{-1}\|(|\bar{g}(s_0)|\|h_{21}\|^2+2\|h_{21}\|+\|h_{21}\|^2\|H_{22}^{-1}\|)\nonumber\\
        &\;\quad\quad +\|H_{22}^{-1}\|\,,\label{eq_Hinv_norm_bd1}
    \end{align}
    Using \eqref{eq_h12_norm_bd}\eqref{eq_H22_norm_bd}\eqref{eq_a_norm_bd}, we can further upper bound \eqref{eq_Hinv_norm_bd1} as
    \be
    \|H^{-1}-\bar{g}(s_0)e_1e_1^\top \|\leq \frac{\lp M_1M_2+1\rp^2}{|f(s_0)|\lambda_2(L)-M_2-M_1M_2^2}\,.\label{eq_Hinv_norm_bd2}
    \ee
    This bound holds as long as $|f(s_0)|\lambda_2(L)>M_2+M_2^2M_1$. Combining \eqref{eq_T_H_norm_equiv} and \eqref{eq_Hinv_norm_bd2} gives the desired inequality.
\end{pf}
\section{Proof of Theorem \ref{thm_to_time_high_con} and \ref{thm_to_time_high_gain}}\label{thm_to_time_pf}
When the input to the network is $U(s)$, the output response of the $i$-th node is $$
        Y_i(s)=e_i^\top T(s)U(s)\,,
    $$
    where $e_i$ is the $i$-th column of the identity matrix $I_n$.
    
    Using Mellin's inverse formula~\cite[Theorem 3.20]{Dullerud2013}, we have
    \begin{align*}
        &\;|y_i(t)-\bar{y}(t)|\\
        =&\;\lvt\frac{1}{2\pi j} \lim_{\omega\ra \infty}\int_{\sigma-j\omega}^{\sigma+j\omega} e^{st}\lp Y_i(s)-e_i^\top \bar{g}(s)\one\frac{\one^\top }{n}U(s)\rp ds\rvt\\
        \leq&\; \frac{e^\sigma}{2\pi}\lim_{\omega\ra \infty}\int_{\sigma-j\omega}^{\sigma+j\omega}\lvt e_i^\top T(s)U(s)-e_i^\top \bar{g}(s)\one\frac{\one^\top }{n}U(s)\rvt ds\\
        \leq &\;\frac{e^\sigma}{2\pi}\lim_{\omega\ra \infty}\int_{\sigma-j\omega}^{\sigma+j\omega} \lV T(s)-\frac{1}{n}\bar{g}(s)\one\one^\top \rV \|U(s)\| ds\\
        =&\;\frac{e^\sigma}{2\pi}\lp (A) + (B) + (C)\rp\,,
    \end{align*}
    where
    $$
        (A)=\int_{\sigma-j\omega_0}^{\sigma+j\omega_0} \lV T(s)-\frac{1}{n}\bar{g}(s)\one\one^\top \rV \|U(s)\|ds\,,
    $$
    $$
        (B)=\lim_{\omega\ra \infty}\int_{\sigma+j\omega_0}^{\sigma+j\omega} \lV T(s)-\frac{1}{n}\bar{g}(s)\one\one^\top \rV \|U(s)\|ds\,,
    $$
    $$
        (C)=\lim_{\omega\ra \infty}\int_{\sigma-j\omega}^{\sigma-j\omega_0} \lV T(s)-\frac{1}{n}\bar{g}(s)\one\one^\top \rV \|U(s)\|ds\,.
    $$
    Both proofs uses such decomposition. By our assumption,
    \begin{align*}
        (B)&=\;\lim_{\omega\ra \infty}\int_{\sigma+j\omega_0}^{\sigma+j\omega} \lV T(s)-\frac{1}{n}\bar{g}(s)\one\one^\top \rV \|U(s)\|ds\\
        &\leq \; \lim_{\omega\ra \infty}\int_{\sigma+j\omega_0}^{\sigma+j\omega} \lp \lV T(s)\rV+\lV\bar{g}(s)\rV \rp\|U(s)\|ds\\
        &\leq\; 2\gamma\lim_{\omega\ra \infty}\int_{\sigma+j\omega_0}^{\sigma+j\omega} \|U(s)\|ds\,,
    \end{align*}
    where the last inequality uses the fact that $\bar{g}(s)$ and $T(s)$ are stable: $\|\bar{g}(s)\|_{\mathcal{H}_\infty},\|T(s)\|_{\mathcal{H}_\infty}\leq \gamma$. Because for the real input signals, we have $U(s^*)=U^*(s)$, hence
    $
        \int^{\sigma-j\omega_0}_{\sigma-j\omega}\|U(s)\|ds=\int^{\sigma+j\omega}_{\sigma+j\omega_0}\|U(s)\|ds\,,
    $
    which leads to
    \ben
        (C)\leq 2\gamma\lim_{\omega\ra \infty}\int_{\sigma+j\omega_0}^{\sigma+j\omega} \|U(s)\|ds\,.
    \een
    Now we are ready to prove Theorem \ref{thm_to_time_high_con} and \ref{thm_to_time_high_gain}.
\begin{pf}[Proof of Theorem \ref{thm_to_time_high_con}]
    First of all, Mellin's inverse formula requires that the vertical line $Re(s)=\sigma$ is on the right of all poles of the signal. This is the case from our assumption that  $\sup_{Re(s)>\sigma}\|U(s)\|<+\infty$ and that $T(s),\bar{g}(s)$ being stable.
    
    By the assumption that $\lim_{\omega\ra \infty}\int_{\sigma+j0}^{\sigma+j\omega} \|U(s)\|ds$ is finite, one can pick an $\omega_0>0$, such that $$\lim_{\omega\ra \infty}\int_{\sigma+j\omega_0}^{\sigma+j\omega} \|U(s)\|ds\leq \frac{2\pi\epsilon}{6e^\sigma\gamma}\,,$$
    which leads to
    \begin{align*}
        (B)&\leq\; 2\gamma\lim_{\omega\ra \infty}\int_{\sigma+j\omega_0}^{\sigma+j\omega} \|U(s)\|ds\leq \frac{2\pi\epsilon}{3e^\sigma }\,.
    \end{align*}
    Similarly, we have $(C)\leq \frac{2\pi\epsilon}{3e^\sigma }$.
    
    For the remaining term, we have
    \begin{align*}
        (A)&=\;\int_{\sigma-j\omega_0}^{\sigma+j\omega_0} \lV T(s)-\frac{1}{n}\bar{g}(s)\one\one^\top \rV \|U(s)\|ds\\
        &\leq\;\sup_{w\in[-w_0,w_0]}\lV T(\sigma+jw)-\frac{1}{n}\bar{g}(\sigma+jw)\one\one^\top \rV\\
        &\;\quad\quad \times\int_{\sigma-j\omega_0}^{\sigma+j\omega_0} \|U(s)\|ds
    \end{align*}
    Since $[\sigma-j\omega_0,\sigma+j\omega_0]$ is a compact set that satisfies the assumption in Theorem \ref{thm_unifm_conv_reg_compact}, we have
    $$
        \lim_{\lambda_2(L)\ra\infty}\sup_{w\in[-w_0,w_0]}\lV T(\sigma+jw)-\frac{1}{n}\bar{g}(\sigma+jw)\one\one^\top \rV=0\,.
    $$
    Therefore, for sufficiently large $\lambda_2(L)$, we have $(A)\leq \frac{2\pi\epsilon}{3e^\sigma}$.
    Combining the upperbounds for $(A),(B),(C)$, we have
    $$
        |y_i(t)-\bar{y}(t)|\leq \epsilon\,.
    $$
    Notice that the choice of $\lambda_2(L)$ does not depends on time $t$, hence this inequality holds for all $t>0$.
\end{pf}
\begin{pf}[Proof of Theorem \ref{thm_to_time_high_gain}]
    Here, the input is a sinusoidal signal $U(s)=\frac{\alpha}{s^2+\alpha^2}u_0,u_0\in\mathbb{S}^{n-1}$. Mellin's inverse formula requires that the vertical line $Re(s)=\sigma$ is on the right of all poles of the signal, which is satisfied under any choice $\sigma>0$. For our purpose, 
    we pick
    \ben
        \sigma=\alpha,\omega_0=K \alpha\,,
    \een
    for some $K>0$ (to be determined later).
    By our assumption,
    \begin{align}
        (B)&\leq\; 2\gamma\lim_{\omega\ra \infty}\int_{\sigma+j\omega_0}^{\sigma+j\omega} \lvt\frac{\alpha}{s^2+\alpha^2}\rvt \|u_0\| ds\nonumber\\
        &=\; 2\gamma\int_{\omega_0}^{+\infty} \frac{\alpha}{|(\sigma+j\omega)^2+\alpha^2|}d\omega\nonumber\\
        &=\; 2\gamma\int_{K\alpha}^{+\infty} \frac{\alpha}{|(\alpha+j\omega)^2+\alpha^2|}d\omega\nonumber\\
        &=\; 2\gamma\int_{K\alpha}^{+\infty} \frac{\alpha}{\sqrt{4\alpha^4+\omega^4}}d\omega\nonumber\\
        &\leq\; 2\sqrt{2}\gamma\int_{K\alpha}^{+\infty}\frac{\alpha}{2\alpha^2+\omega^2}d\omega\nonumber\\
        &=\;\gamma\lp \pi-2\arctan\lp\frac{K}{\sqrt{2}}\rp\rp\,,\label{eq_pf_thm_to_time_high_gain_1}
    \end{align}
    where the last inequality use the fact that for $a,b>0$, we have 
    \ben
    \sqrt{a^2+b^2}\geq (a+b)/\sqrt{2}\,.
    \een
    Similarly, we have
    \be
        (C)\leq \gamma\lp \pi-2\arctan\lp\frac{K}{\sqrt{2}}\rp\rp\,.\label{eq_pf_thm_to_time_high_gain_2}
    \ee
    For the remaining term, we use the result in the proof of Theorem \ref{thm_ptw_singular_f_pole}: $\exists \delta>0$, such that $\forall s\in \mathcal{B}(0,\delta)$ such that
    \ben
        \lV T(s)-\frac{1}{n}\bar{g}(s)\one\one^\top \rV\leq 
        \frac{2\lp M_1M_2+1\rp^2}{|f(s)|\lambda_2(L)}\,,
    \een
    for some $M_1,M_2>0$. Then as long as we pick $\alpha,K$ appropriately such that $|\sigma+j\omega_0|\leq \delta$, i.e., $\sqrt{1+K^2}\alpha\leq \delta$, we have
    \begin{align*}
        (A)&=\;\int_{\sigma-j\omega_0}^{\sigma+j\omega_0} \lV T(s)-\frac{1}{n}\bar{g}(s)\one\one^\top \rV \lvt\frac{\alpha}{s^2+\alpha^2}\rvt ds\\
        &\leq\;\int_{\sigma-j\omega_0}^{\sigma+j\omega_0}\frac{2\lp M_1M_2+1\rp^2}{|f(s)|\lambda_2(L)}  \lvt\frac{\alpha}{s^2+\alpha^2}\rvt ds\\
        &=\;\int_{\sigma-j\omega_0}^{\sigma+j\omega_0}\frac{2\lp M_1M_2+1\rp^2}{\lambda_2(L)/|s|}  \frac{\alpha}{|s^2+\alpha^2|} ds\\
        &=\frac{2\lp M_1M_2+1\rp^2}{\lambda_2(L)} \;\int_{\sigma-j\omega_0}^{\sigma+j\omega_0} \frac{|s|\alpha}{|s^2+\alpha^2|} ds\\
        &=\frac{4\lp M_1M_2+1\rp^2}{\lambda_2(L)} \;\int_{0}^{K\alpha} \frac{|\alpha+j\omega|\alpha}{|(\alpha+j\omega)^2+\alpha^2|}d\omega\\
        &=\frac{4\lp M_1M_2+1\rp^2}{\lambda_2(L)} \;\int_{0}^{K\alpha} \frac{\sqrt{\alpha^2+\omega^2}\alpha}{\sqrt{4\alpha^4+\omega^4}}d\omega\\
        &\leq\; \frac{2\sqrt{2}\lp M_1M_2+1\rp^2}{\lambda_2(L)} \;\int_{0}^{K\alpha} \frac{2(\alpha+\omega)\alpha}{2\alpha^2+\omega^2}d\omega\,,
    \end{align*}
    where the last equality used the fact that for $a,b>0$, we have 
    \ben
    a+b\geq \sqrt{a^2+b^2}\geq (a+b)/\sqrt{2}\,,
    \een
    to upper and lower bound the numerator and denominator respectively. Notice that 
    \begin{align}
        &\;\int_{0}^{K\alpha} \frac{2(\alpha+\omega)\alpha}{2\alpha^2+\omega^2}d\omega\nonumber\\
        &=\;\alpha\lp \sqrt{2}\arctan\lp\frac{K}{\sqrt{2}}\rp+\log\lp 1+\frac{K^2}{2}\rp\rp\nonumber\\
        &\leq\; 2\alpha \log\lp\frac{K^2}{2}\rp\,,\label{eq_pf_to_time_high_gain_1}
    \end{align}
    for sufficiently large $K$. We have
    \be
        (A)\leq \frac{4\sqrt{2}\lp M_1M_2+1\rp^2}{\lambda_2(L)}\alpha \log\lp \frac{K^2}{2}\rp\,.\label{eq_pf_thm_to_time_high_gain_3}
    \ee
    The last step is to find the right choice of $\alpha,K$. Given $\epsilon>0$, pick a $K>0$, such that \ben
        2\gamma\lp \pi-2\arctan\lp\frac{K}{\sqrt{2}}\rp\rp\leq \frac{\epsilon \pi}{2}\,.
    \een
    Generally such a $K$ is sufficient for \eqref{eq_pf_to_time_high_gain_1} to hold.
    With this choice of $K$, let 
    \begin{align*}
        &\;\alpha_0:=\\
        &\;\min\lb \log 2, \frac{\epsilon \pi \lambda_2(L)}{8\sqrt{2}(M_1M_2+1)^2\log\lp \frac{K^2}{2}\rp},\frac{\delta}{\sqrt{1+K^2}}\rb\,.
    \end{align*}
    Then, $\forall \alpha\leq \alpha_0$, combining \eqref{eq_pf_thm_to_time_high_gain_1}\eqref{eq_pf_thm_to_time_high_gain_2}\eqref{eq_pf_thm_to_time_high_gain_3}, we have
    \begin{align*}
        &\;|y_i(t)-\bar{y}(t)|\\
        &\;\leq \frac{e^\sigma}{2\pi}((A)+(B)+(C))\\
        &\;\leq \frac{e^{\alpha_0}}{2\pi}\lp 2\gamma\lp \pi-2\arctan\lp\frac{K}{\sqrt{2}}\rp\rp \right.+ \\
        &\; \qquad\left. \frac{4\sqrt{2}\lp M_1M_2+1\rp^2}{\lambda_2(L)}\alpha \log\lp \frac{K^2}{2}\rp \rp\\
        &\leq\; \frac{1}{\pi} \lp \frac{\epsilon \pi}{2}+\frac{\epsilon \pi}{2}\rp=\epsilon\,.
    \end{align*}
    Notice that the choice of $\alpha_0,K$ does not depends on time $t$, nor the node index $i$, hence this inequality holds for all $t>0$ and all $i\in[n]$.
\end{pf}
\section{Proof of Theorem \ref{thm_passive_to_stable}}\label{thm_to_passive_to_stable_pf}
For each $g_i(s), i=1,\cdots,n$, we have, by the OSP property,
    $$
        Re(g_i(s))\geq \epsilon |g_i(s)|^2,\forall Re(s)>0\,.
    $$
    That is,
    $$
        Re(G(s))\succeq \epsilon  G^*(s)G(s)\,, 
    $$
    or equivalently,
    $
        \bmt G(s) \\ I\emt^* \bmt -\epsilon I & I\\ I & 0\emt\bmt G(s) \\ I\emt\succeq 0\,.
    $
    Since $g_i(s)$ are all OSP, then $g_i(s)$ is positive real~\cite{khalil2002nonlinear}. A positive real function that is not zero function has no zero nor pole on the left half plane. Therefore $g_i(s)$ are invertible for all $Re(s)>0$, which ensures that $G(s)$ is invertible for all $Re(s)>0$.
    Then 
    $$
        (G^*(s))^{-1}\bmt G(s) \\ I\emt^* \bmt -\epsilon I & I\\ I & 0\emt\bmt G(s) \\ I\emt G^{-1}(s)\succeq 0\,,
    $$
    which is
    \be
        \bmt I \\ G^{-1}(s)\emt^* \bmt -\epsilon I & I\\ I & 0\emt\bmt I \\ G^{-1}(s)\emt\succeq 0\,.\label{eq_osp_G}
    \ee
    Notice that $$T(s)=(I+G(s)f(s)L)^{-1}G(s)=(G^{-1}(s)+f(s)L)^{-1}\,,$$
    then from \eqref{eq_osp_G} and the fact that $f(s)$ is PR, we have
    \begin{align*}
        &\;\bmt I \\ T^{-1}(s)\emt^* \bmt -\epsilon I & I\\ I & 0\emt\bmt I \\ T^{-1}(s)\emt\\
        =&\;\bmt I \\ G^{-1}(s)+f(s)L\emt^* \bmt -\epsilon I & I\\ I & 0\emt\bmt I \\ G^{-1}(s)+f(s)L\emt\\
        =&\; \bmt I \\ G^{-1}(s)\emt^* \bmt -\epsilon I & I\\ I & 0\emt\bmt I \\ G^{-1}(s)\emt+[f^*(s)+f(s)]L\\
        \succeq&\; \bmt I \\ G^{-1}(s)\emt^* \bmt -\epsilon I & I\\ I & 0\emt\bmt I \\ G^{-1}(s)\emt\succeq 0\,.
    \end{align*}
    Now for sufficiently large $\gamma>0$, we have
    $$
        \bmt -\epsilon I & I\\ I & 0\emt+\bmt \frac{\epsilon}{2} I & 0\\ 0 & -\gamma^2 \frac{\epsilon}{2} I\emt=\bmt -\frac{\epsilon}{2} I & I\\ I & -\gamma^2 \frac{\epsilon}{2} I\emt\preceq 0\,,
    $$
    since its Schur complement $(-\frac{\epsilon}{2}+\frac{2}{\epsilon \gamma^2})I\preceq 0$ for large $\gamma$.
    Therefore,
    \begin{align*}
        &\;\bmt I \\ T^{-1}(s)\emt^* \bmt -\frac{\epsilon}{2} I & 0\\ 0 & \gamma^2 \frac{\epsilon}{2} I\emt\bmt I \\ T^{-1}(s)\emt\\
        \succeq&\;\bmt I \\ T^{-1}(s)\emt^* \bmt -\epsilon I & I\\ I & 0\emt\bmt I \\ T^{-1}(s)\emt\succeq 0\,,
    \end{align*}
    which is exactly,
    $\gamma^2\frac{\epsilon}{2}(T^{-1}(s))^*(T^{-1}(s))\succeq \frac{\epsilon}{2}I\,.$
    This shows that
    \be
        \sigma_{min}^2(T^{-1}(s))\geq \frac{1}{\gamma^2},\forall Re(s)>0\,,\label{eq_time_passive_pf_1}
    \ee
    which is equivalent to
    $
        \|T(s)\|_2\leq \gamma\,, \forall Re(s)>0\,.
    $
    Moreover, \eqref{eq_time_passive_pf_1} implies
    $$ 
    |\bar{g}^{-1}(s)|=\lvt\frac{\one^\top }{\sqrt{n}} T^{-1}(s)\frac{\one}{\sqrt{n}}\rvt^2\geq \frac{1}{\gamma^2},\forall Re(s)>0\,,
    $$
    which is equivalent to $\|\bar{g}(s)\|_2\leq \gamma,\forall Re(s)>0$.

\ifthenelse{\boolean{archive}}{\section{Additional Convergence Results of $T(s)$}
    Before presenting with the results, we define
    \begin{defn}
        For transfer function $g(s)$ and $s_0\in \compl$, $s_0$ is a generic point of $g(s)$ if $s_0$ is neither a pole nor a zero of $g(s)$.
    \end{defn}
    
    \subsection{Convergence at Generic Points of $f(s)$}\label{ssec:gen_p_of_f}
    In this section we keep $s_0$ fixed and present the point-wise convergence result of $T(s_0)$ as $\lambda_2(L)$ increases. This requires $s_0$ to be a generic point of $f(s)$.

    A careful analysis shows that, as $\lambda_2(L)\ra +\infty$, the pole of $\bar{g}(s)$ is asymptotically a pole of $T(s)$, and the zero of $\bar{g}(s)$ is asymptotically a zero of $T(s)$, as stated in the following two theorems.
    \input{old_files/subfiles/thm_ptw_conv_pole/statement.tex}
    \begin{rem}\label{rem_pole}
        Theorem \ref{thm_ptw_conv_pole} does not suggest whether the network is asymptotically coherent at poles of $\bar{g}(s)$. Our incoherence measure $\lV T(s_0)-\frac{1}{n}\bar{g}(s_0)\one\one^T\rV$ is undefined at such poles. Alternatively, for $s_0$ the pole of $\bar{g}(s)$, one can prove that when $\tilde{\Lambda}/\tilde{\lambda_2(L)}\ra \Lambda_{\mathrm{lim}}$ as $\lambda_2(L)\ra +\infty$, we have the limit $\lV \frac{T(s_0)}{\|T(s_0)\|}-\frac{1}{n}\gamma(\Lambda_\mathrm{lim})\one\one^T\rV\ra 0$, for some $\gamma(\Lambda_\mathrm{lim})\in\compl$ determined by $\Lambda_\mathrm{lim}$ with $|\gamma(\Lambda_\mathrm{lim})|=1$. This suggests that $T(s_0)$ has the desired rank-one structure for coherence. While the normalized transfer matrix is not discussed in this paper due to the space constraints, such formulation is better for understanding the network coherence at pole of $\bar{g}(s)$.
    \end{rem}
    Next, the convergence result regarding the zeros of $\bar{g}(s)$ is stated as
    \input{old_files/subfiles/thm_ptw_conv_zero/statement.tex}
    \begin{rem}\label{rem_zero}
        The limit in Theorem \ref{thm_ptw_conv_zero} can still be written as $\lim_{\lambda_2(L)\ra +\infty}\|T(s_0)-\frac{1}{n}\bar{g}(s_0)\one\one^T\|=0$, because $s_0$ is a zero of $\bar{g}(s)$. However, we here emphasize the fact that the system is not coherent  at $s_0$ under normalization because $T(s_0)/\|T(s_0)\|$ does not converge to $\frac{1}{n}\gamma\one\one^T$ for any $\gamma\in\compl$. 
    \end{rem}
    So far, we have shown point-wise convergence of $T(s)$ towards the transfer function $\frac{1}{n}\bar{g}(s)\one\one^T$, 
    from which we assess how network coherence emerges as connectivity increases. In Remark \ref{rem_pole} and \ref{rem_zero} we see the incoherence measure $\lV T(s_0)-\frac{1}{n}\bar{g}(s_0)\one\one^T\rV$ is insufficient for understanding the asymptotic behavior at zero and poles of $\bar{g}(s)$, and the alternative measure $\lV \frac{T(s_0)}{\|T(s_0)\|}-\frac{1}{n}\gamma\one\one^T\rV$ is a good complement\footnote{As $\lambda_2(L)$ increases, for pole of $\bar{g}(s)$, the latter converges to $0$ given suitable conditions but not for the former; for zero of $\bar{g}(s)$, the opposite result holds; for generic point of $\bar{g}(s)$, both incoherence measures converge to $0$.} for such purpose. The latter focus more on the relative scale of eigenvalues of $T(s)$. In this paper, we mostly use the former, $\lV T(s_0)-\frac{1}{n}\bar{g}(s_0)\one\one^T\rV$, because we're interested in connecting these results to the network time-domain response.
    
    
    
    \subsection{Uniform Convergence Around Zeros of $\bar{g}(s)$}
    We have seen in Theorem \ref{thm_unifm_conv_reg_compact} that $T(s)$ can have a low-rank structure over a compact set $S$ containing no zero nor pole of $\bar{g}(s)$. We consider here the case $S$ contains zero of $\bar{g}(s)$.
    
    Similarly to the point-wise convergence, we discuss uniform convergence of $T(s)$ over set $S$ that satisfies the following assumption
    \begin{assum}
        $S\subset\compl$ satisfies $\sup_{s\in S}|f(s)|<\infty$ and $\inf_{s\in S}|f(s)|>0$.\label{asmp_unifm}
    \end{assum}
    Such an assumption guarantees all points in the closure of $S$ are generic points of $f(s)$. This property prevents any sequence of points in $S$ that asymptotically eliminates or amplifies the network effect on the boundary of $S$. In subsequent sections, we denote $F_h:=\sup_{s\in S}|f(s)|$ and $F_l:=\inf_{s\in S}|f(s)|>0$. 
    
    We first define the notion of \emph{Nodal Multiplicity} of a point in complex plane w.r.t. a given network.
    \input{old_files/subfiles/def_nodal_multi/statement.tex}
    By definition, any zero of $\bar{g}(s)$ must have positive nodal multiplicity. Our finding is that zeros with nodal multiplicity exactly $1$ have a special property, which is shown in the following Lemma.
    \input{old_files/subfiles/lem_unifm_conv_zero_neighbor/statement.tex}
    The proof is shown in Appendix \ref{app_proof_lem_unifm_conv_zero_neighbor}. Notice that for given $\epsilon>0$, the $\epsilon$ bound is valid for any $\lambda_2(L)\geq \lambda$, therefore we can prove uniform convergence over compact regions that only contain zeros of $\bar{g}(s)$ with nodal multiplicity $1$.
    \input{old_files/subfiles/thm_unifm_conv_compact/statement.tex}

    For zeros with nodal multiplicity strictly larger than $1$, the analysis is rather complicated. We first look once again at the homogeneous node dynamics setting of Section \ref{sec:prem} to provide some insight. 
    
    \begin{exmp}
    Consider again a homogeneous network with node dynamics $g(s)$ and $f(s)=1$, where the transfer matrix is given by
    \ben
        T(s) = \frac{1}{n}g(s)\one\one^T+V_\perp\dg\lb \frac{1}{g^{-1}(s)+\lambda_i(L)}\rb_{i=2}^n V_\perp^T\,.
    \een
    The poles of $T(s)$ include 1) the poles of $g(s)$, and 2) any point $s_0$ that satisfies $g^{-1}(s_0)+\lambda_i(L)=0$ for a particular $i$. Notice that if $\lambda_i(L)$ is large, every solution to $g^{-1}(s_0)+\lambda_i(L)=0$ is close to one of the zeros of $g(s)$. As we increase $\lambda_2(L)$, which effectively increases every $\lambda_i(L),2\leq i\leq n$, one can check that at most $n-1$ poles asymptotically approach each zero of $g(s)$, provided that $\lambda_i(L)$ are distinct. As a result, uniform convergence around any zero of $g(s)$ cannot be obtained due to the presence of poles of $T(s)$ close to them.
    \end{exmp}
    
    Such observation also seems to hold in general for networks with heterogeneous node dynamics $g_i(s)$. That is, if a zero of $\bar{g}(s)$ is a zero with nodal multiplicity strictly larger than $1$, then we expect it to ``attract" poles of $T(s)$. But it is difficult to formally prove it since we cannot exactly locate the poles of $T(s)$ in the absence of homogeneity.\footnote{We can still exactly locate the poles of $T(s)$ when proportionality is assumed, i.e. $g_i(s)=f_ig(s),i\in[n]$ for some $f_i>0$ and rational transfer function $g(s)$. Such a case can be regarded as the homogeneous case by considering a scaled version of $L$.} Surprisingly, there are certain cases where we can still quantify the effect of those poles of $T(s)$ approaching a zero of $\bar g(s)$. This essentially disproves the uniform convergence for such cases.
    \input{old_files/subfiles/thm_unifm_conv_fail/statement.tex}
    \ifthenelse{\boolean{archive}}{The proof is shown in Appendix \ref{app_proof_thm_unifm_conv_fail}}{The proof considers the first-order taylor approximation of $G(s)$ around the zero $z$ and is mostly technical. Due to the space constraint, we refer interested readers to~\cite{??} for the proof}. Although Theorem \ref{thm_unifm_conv_fail} only disproves the uniform convergence around one particular type of zero of $\bar{g}(s)$, namely, such zero must have nodal multiplicity $n$ and it must be a real zero with multiplicity 1 for all $g_i(s)$, we believe a similar result holds for any zero that is ``shared" by multiple $g_i(s)$. However, a complete proof is left for future research. 
    
    We now provide another point of view of this phenomenon. Suppose in a network of size 2, $z_1$ is exclusively zero of $g_1(s)$ and $z_2$ exclusively zero of $g_2(s)$. If $z_1$ and $z_2$ are close enough, there must be $p$ a pole of $\bar{g}(s)$ in the small neighborhood of $z_1$ or $z_2$. To be more clear, see the following example.
    \begin{exmp}
        Let $g_1(s)=\frac{s+a}{s^2},g_2(s)=\frac{s+a+\epsilon}{s^2}$, then $z_1=-a$ and $z_2=-a-\epsilon$ are the zeros respectively. The coherent dynamics is given by
        \ben
            \bar{g}(s)=\frac{2}{g_1^{-1}(s)+g_2^{-1}(s)}=\frac{(s+a)(s+a+\epsilon)}{2s^2(s+a+\epsilon/2)}\,.
        \een
        $\bar{g}(s)$ has a pole $p=-a-\epsilon/2$ that is in both $\epsilon/2$-neighborhoods of $z_1$ and $z_2$.
    \end{exmp}
    By Theorem \ref{thm_ptw_conv_pole}, we know that $p$ is asymptotically a pole of $T(s)$, in other words, there is a pole of $T(s)$ approaching $p$, as the network connectivity increases. Moreover, $z_1$ and $z_2$ being close enough suggests that $p$ is close to $z_1$ and $z_2$, as we see in the example. Consequently, two zeros $z_1,z_2$ being close introduces a pole of $T(s)$ asymptotically approaching a small neighborhood of $z_1,z_2$. Consider the limit case where the two zeros collapse into a shared zero of $g_1(s),g_2(s)$, we should expect a pole of $T(s)$ approaching this shared zero.
    
    A similar argument can be made for $m$ zeros of different nodes being close to each other, introducing $m-1$ poles of $\bar{g}(s)$ in the small neighborhood that asymptotically attract poles of $T(s)$. This is by no means a rigorous proof of how uniform convergence fails around a zero ``shared" by multiple $g_i(s)$, but rather a discussion providing  intuition behind such behavior.
    
    At this point, we have proved uniform convergence of $T(s)$ on a compact set $S$ that does not include 1) zeros of $\bar{g}(s)$ with Nodal Multiplicity larger than $1$, or 2) poles of $\bar{g}(s)$.
    
    In particular, we find that zero with Nodal Multiplicity larger than $1$, i.e. it is "shared" by multiple $g_i(s)$, attracts pole of $T(s)$ as network connectivity increases, which suggests that uniform convergence of $T(s)$ fails around such point. Although we only provide the proof for special cases as in Theorem \ref{thm_unifm_conv_fail}, we conjecture such a statement is true in general and we left more careful analysis for future research. 
    
    \subsection{Uniform Convergence on Right-Half Complex Plane}
    Aside from uniform convergence on compact sets, uniform convergence over the closed right-half plane $\{s:Re(s)\geq 0\}$ is of great interest as well. If we were to establish uniform convergence over the right-half plane for a certain $T(s)$, then given $\bar{g}(s)$ to be stable, the convergence in $\mathcal{H}_\infty$-norm of $T(s)$ towards $\frac{1}{n}\bar{g}(s)\one\one^T$ could be guaranteed, i.e., $T(s)$ converges to $\frac{1}{n}\bar{g}(s)\one\one^T$ as a system. One trivial consequence is that we can infer the stability of $T(s)$ with a large enough $\lambda_2(L)$ by the stability of $\bar{g}(s)$. Furthermore, given any $\mathcal{L}_2$ input signal, we can make the $\mathcal{L}_2$ difference between output responses of $T(s)$ and $\frac{1}{n}\bar{g}(s)\one\one^T$ arbitrarily small by increasing the network connectivity. 
    
    Unfortunately, for most networks, we encounter with the same issue we have seen when dealing with zeros of $\bar{g}(s)$: When $g_i(s)$ is strictly proper, $g_i(s)\ra 0$ as $|s|\ra +\infty$, thus, $\infty$ can be viewed as a zero of $g_i(s)$ by regarding $g_i(s)$ as functions defined on extended complex plane $\compl \cup \{\infty\}$. Then for networks that include more than one node whose transfer functions are strictly proper, there will be poles of $T(s)$ approaching $\{\infty\}$ as $\lambda_2(L)$ increases. Notice that those poles could approach $\{\infty\}$ either from the left-half or right-half plane. Apparently, the uniform convergence on the right-half plane will not hold if the latter happens, but even when the former happens, we still need to quantify the effect of such poles because they are approaching the boundary of our set $\{s:Re(s)\geq 0\}$. A similar argument can be made for any set of the form $\{s:Re(s)=\sigma\}=\{\sigma+j\omega:\omega\in[-\infty,+\infty]\}$, which we mentioned in \ref{ssec:goal}.
    
    Although proving (or disproving) uniform convergence on the right-half plane for general networks is quite challenging, it is much more straightforward for networks consist of only non-strictly proper nodes, as shown in the following theorem:  
    \input{old_files/subfiles/thm_unifm_conv_rh/statement.tex}
}{}
\bibliographystyle{ieeetr}        
\bibliography{ref.bib}           

\end{document}